\documentclass[USenglish,oneside,twocolumn]{article}

\usepackage{graphicx} 
\usepackage{etex}
\usepackage[big]{dgruyter_NEW}
\usepackage{enumerate}
\usepackage{epsfig}

\usepackage{amsmath}
\usepackage{amssymb}
\usepackage{algorithm}
\usepackage{url}
\usepackage{listings}
\usepackage{subfigure}
\usepackage{amsmath}
\usepackage{rotating}
\usepackage{lscape}
\usepackage{verbatim}
\usepackage{hhline}
\usepackage{textcomp,booktabs}
\usepackage{xcolor}
\usepackage{paralist}
\usepackage{textcomp}
\usepackage{hyperref}
\usepackage{fixltx2e}

\usepackage{numprint}
\npthousandsep{,}

\usepackage{fp}
\cclogo{\includegraphics{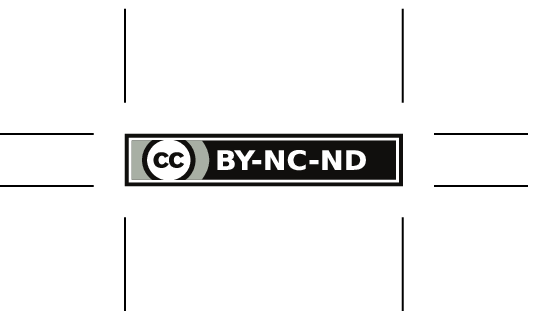}}

\usepackage{url}

\newcommand{\btcPrice}{6080} 
\newcommand{\btcFee}{9} 
\newcommand{\ptwoshPrice}[1]{\FPeval{\result}{round(#1*\btcFee,0)}}
\newcommand{\usd}[1]{\numprint {#1}}

\newcommand{\satToUSD}[1]{\FPeval{\result}{round(#1*\btcPrice/100000000,2)}}

\begin{document}

\author*[1]{Ruben Recabarren}
\author[2]{Bogdan Carbunar}
\affil[1]{Florida Int'l University, Miami, FL 33199, E-mail: recabarren@gmail.com}
\affil[2]{Florida Int'l University, Miami, FL 33199, E-mail: carbunar@gmail.com}

\title{\huge{Tithonus: A Bitcoin Based Censorship Resilient System}}

\runningtitle{Tithonus: A Bitcoin Based Censorship Resilient System}

\begin{abstract}
{Providing reliable and surreptitious communications is difficult in the
presence of adaptive and resourceful state level censors. In this paper we
introduce Tithonus, a framework that builds on the Bitcoin blockchain and
network to provide censorship-resistant communication mechanisms. 
In contrast to previous approaches, we do not rely solely on the
slow and expensive blockchain consensus mechanism but instead fully exploit
Bitcoin's peer-to-peer gossip protocol. We develop adaptive, fast and cost
effective data communication solutions that camouflage client requests into
inconspicuous Bitcoin transactions. We propose solutions to securely request
and transfer content, with unobservability and censorship resistance, and free,
pay-per-access and subscription based payment options.  When compared to
state-of-the-art Bitcoin writing solutions, Tithonus reduces the cost of
transferring data to censored clients by 2 orders of magnitude and increases
the goodput by 3 to 5 orders of magnitude. We show that Tithonus client
initiated transactions are hard to detect, while server initiated transactions
cannot be censored without creating split world problems to the Bitcoin
blockchain.
}
\end{abstract}

\journalname{Proceedings on Privacy Enhancing Technologies}
\DOI{Editor to enter DOI}
\startpage{1}
\received{..}
\revised{..}
\accepted{..}

\journalyear{..}
\journalvolume{..}
\journalissue{..}

\maketitle

\section{Introduction}

Evading Internet censorship is difficult in the presence of state level censors
that continuously adapt and employ state of the art
technologies~\cite{MSRMBP14,BlueCoat,Cisco,FinFisher,HackingTeam} to e.g.,
block IP addresses, perform deep packet inspection, and corrupt protocols
including BGP hijacking~\cite{Iran} and DNS manipulation~\cite{PJLEFWP17}. In
this paper we propose a new approach to bypass such censors, that exploits the
distributed resilience of the Bitcoin cryptocurrency, and the substantial
collateral damage~\cite{HH15,ZH16,FLHWP15} inflicted by blocking access to or
tampering with its blockchain and network.

We introduce Tithonus\footnote{Titan who was granted eternal life but not
eternal youth.}, a framework that leverages Bitcoin transactions to provide
unobservable and censorship-resistant communications, and financially reward
participation.

We survey transaction-based Bitcoin blockchain writing mechanisms, and show
that their high cost and low bitrate make them unsuitable for censorship
resilient communications. Instead, we observe that transactions do not need to
reach the blockchain in order to communicate data, and leverage the Bitcoin
network to propagate and receive transactions in seconds. We reduce the fees
paid in such {\it swift transactions}, to the level where they are valid and
unobservable by censors. To further improve the goodput of Tithonus
communications and reduce their observability, we document, correct, extend and
implement {\it staged transactions}.

We introduce new techniques to transmit messages of arbitrary size written into
multiple Tithonus transactions, that preserve their sequence and
unobservability. Unlike previous solutions like Catena~\cite{TD17} and
Blockstack~\cite{ANSF16} that use the immutability and unequivocation of the
blockchain to persist data, we delegate security to a higher layer and use the
blockchain in its most insecure form. This enables our proposed {\it hidden
sequencing} techniques to achieve significantly lower costs and higher goodput
than state of the art solutions, and prevent a monitoring censor from learning
relationships between data units of the same message.

Tithonus provides mechanisms to bootstrap trust and establish secure channels,
and enables clients to access both static and dynamic censored content with
flexible payment options. Tithonus minimizes the number and size of client
initiated messages, and camouflages them into popular Bitcoin transactions.

Further, we propose an {\it altruistic directory} approach, where clients find and
download blockchain-persisted, static content for free, with unobservability,
uncensorability and integrity assurances. We introduce an {\it on-demand,
pay-per-access} solution, where clients securely request and pay for new
content, with unobservability and censorship resilience, while the Tithonus
server communicates and caches content, with uncensorability and integrity
assurances.

The Bitcoin infrastructure used for Tithonus communications renders ineffective
censorship that uses IP blocking and network protocol manipulation. We reveal
however that the Bitcoin network has a significant number of non-conformant nodes
in countries with known censorship practices. We show that Tithonus is robust
to such nodes, and this robustness is not affected by the use of low fee
transactions, in the absence of congestion. As a consequence, Tithonus is able
to provide an optimally cheap solution within a given cryptocurrency ecosystem.
Furthermore, our experiments show that Tithonus is practical when considering
the reach and number of Bitcoin nodes available in censored countries.

The use of payments enables communications through the Bitcoin ecosystem,
provides incentives for the Tithonus service operation, and
prevents DoS attacks.  Our evaluation is done on the cryptocurrency with the
highest market share, thus the hardest to censor but also the most expensive.
In summary, we introduce the following contributions:

\begin{compactitem}

\item
{\bf Communications over Bitcoin's gossip protocol}.
We are the first to propose the use of Bitcoin's gossip protocol as a direct
medium to exchange arbitrary information instead of relying on the slow and
more expensive blockchain consensus mechanism.

\item
{\bf Bitcoin based censorship circumvention}.
We are the first to leverage the collateral damage inflicted by blocking
Bitcoin, and the intrinsic censorship resistance of its blockchain and
network, to develop censorship resilient communication solutions.

\item
{\bf Tithonus}.
We develop secure, fast and cost effective solutions to communicate data
between censored areas and the free world with censorship resistance and
unobservability. We devise techniques to embed arbitrary encrypted
data indistinguishable from public keys in Bitcoin transactions.

\item
{\bf Prototype implementation}.
We implemented Tithonus infrastructure components in Python. When compared to state-of-the-art Blockchain writing solutions, 
Tithonus reduces the cost of sending data to Bitcoin nodes
in censored countries, by 2 orders of magnitude, and increases the writing
efficiency 3 fold, resulting in a goodput increase of between 3-5 orders of
magnitude. When using the lowest fee, 1/8 of the nodes in censored countries
relayed Tithonus transactions in under 5s.

\end{compactitem}

\section{Background}
\label{sec:background}

\noindent
{\bf Bitcoin transactions}.
We briefly describe the components of a Bitcoin transaction, some of which are
used to store Tithonus data, see Section\ref{sec:background:writing}. A Bitcoin
transaction (see Figure~\ref{fig:txn:p2sh} for an illustration) consists of a
series of inputs and outputs that follow a set of rules.
An input consists of (1) a {\it pointer to a previous transaction} that
contains a funding output, (2) an {\it offset} pointing to the specific output
on the funding transaction, and (3) a \textit{script} (called the
\textit{scriptSig} script) used to verify that a user is authorized to spend
the balance.
An output consists of (1) a {\it value} that is to be transferred from the sum
of values specified on the list of inputs and (2) a {\it script} (called the
\textit{scriptPubKey} script) that specifies how to claim the transferred value
in future transactions.
The transaction is invalid if the sum of the values from the inputs is smaller
than or equal to the sum of the outputs. The balance after subtracting the
output values is considered to be the {\it miner fee}.

\vspace{-5pt}

\subsection{Blockchain Writing}
\label{sec:background:writing}

\begin{figure}[t]
\centering
\includegraphics[width=1\columnwidth]{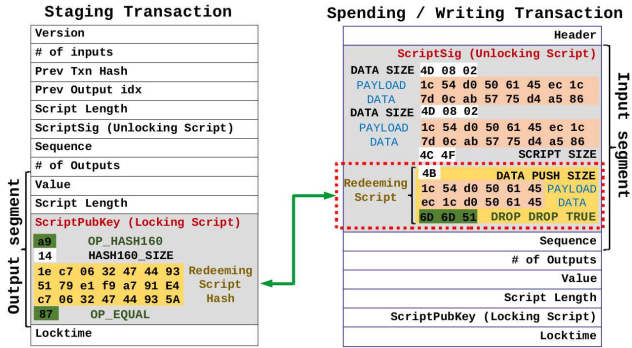}
\caption{Pay-To-Script-Hash (p2sh) transaction with {\it scriptSig} overloading
(1,635 bytes). A \textit{preparing} transaction prepares the spending of the
\textit{redeeming} transaction, that overloads \textit{scriptSig}.}
\label{fig:txn:p2sh}
\vspace{-15pt}
\end{figure}

We survey relevant solutions for writing in Bitcoin transactions, which are
persisted into the blockchain.

\noindent
{\bf Overwriting destination addresses}.
One of the first studied blockchain writing solutions~\cite{kaminskyPayToHash}
used the output address bytes in the {\it scriptPubKey} to store arbitrary
data. Two types of contracts are most commonly used for this: Pay-to-PubkeyHash
(p2pkh) and Pay-to-Script-Hash (p2sh). The p2pkh writing method (used by e.g.,
Apertus~\cite{Apertus}) uses the 20 bytes of a destination address to store
arbitrary data.
Similarly, the output overloading of a p2sh transaction allows for the
overwriting of the redeeming script hash (20 bytes)
Since this is arbitrary data, redeeming this output becomes highly unlikely.

\begin{figure}[t]
\centering
\includegraphics[width=0.89\columnwidth]{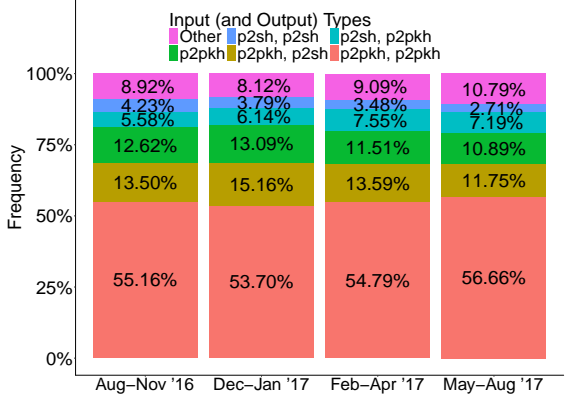}
\vspace{-5pt}
\caption{Input (and output) type distribution between August 2016 and August
2017. Tithonus leverages the observation that p2pkh and p2sh inputs and outputs
are the most frequent, to generate transactions that do not stand out.}
\label{fig:output:distrib}
\vspace{-10pt}
\end{figure}

\noindent
{\bf Overwriting destination public keys}.
A similar technique overwrites public keys instead of destination addresses,
using an old Bitcoin contract, the Pay-To-PubKey (not the Pay-To-PubKey Hash).
While this script is now considered obsolete, it is still valid and accepted by
some miners. Depending on the use of compressed or uncompressed public keys,
this technique can insert 33 or 65 bytes of effective payload data.

\noindent
{\bf OP\_RETURN}.
This writing technique, used by Blockstack~\cite{ANSF16} and Catena~\cite{TD17}
to prevent equivocation, requires the use of an OP\_RETURN opcode in a
scriptPubKey script, which marks the transaction as invalid. Thus, its outputs
are un-spendable, and immediately prunable from the Bitcoin un-spent
transaction set. Officially, this contract allows for writing 80 bytes after
the OP\_RETURN opcode.

\noindent
{\bf Overwriting input scripts}.
A writing technique which we call {\it staged transactions}, has been
documented~\cite{inputWritingPeterTodd}, to exploit the large script in p2sh
inputs. It consists of two transactions, see Figure~\ref{fig:txn:p2sh}. The
first, ``staging'' transaction has a p2sh output that specifies the hash of its redeeming script. In a second, ``writing'' transaction, the input script
provides a {\it redeemScript} that satisfies the staging transaction's conditions. The redeeming script along with the whole \textit{scriptSig}
can then be used to store data. Figure~\ref{fig:txn:p2sh} shows the
use of a construct that simply pushes data to the virtual machine stack.
However, other constructions are possible including using a MULTISIG script,
that allows for any $m$ out of $n$ signatures to claim the corresponding
balance.
The outputs of the writing transaction can be p2pkh, OP\_RETURN or another p2sh
(see above).

\noindent
{\bf Input and output type distribution}.
Figure~\ref{fig:output:distrib} shows the distribution of the input types
(identical to the distribution of output types) in transactions mined between
August 2016 and August 2017. p2pkh inputs and outputs are the most frequent,
especially in transactions with 2 p2pkh inputs or outputs. p2sh inputs and
outputs are the next most popular ones. Thus, in order to embed data into
transactions that are indistinguishable from regular Bitcoin transactions, a
solution needs to use a mix of such inputs and outputs.

For a more detailed discussion on these writing techniques see for instance
Sward et al. \cite{sward2018data}.

\vspace{-5pt}

\subsection{Bitcoin Network}
\label{sec:background:network}

The backbone of the Bitcoin network interconnects participating devices using a
TCP/IP based protocol. Each node
uses hardcoded rules to find other nodes, create outgoing connections to and
optionally accept incoming connections from peers that participate in the
network. Peers opportunistically exchange information about the Bitcoin system
using specific messages.

The \textit{inv} (inventory) message, is used to communicate information (i.e.,
hashes) about all new transactions and blocks in the network. A receiving full
node needs to request further information for all previously unseen objects
received from peers, by issuing a \textit{getdata} message.
However, nodes (e.g., light nodes) that are not equipped with enough resources
to handle large volumes of object information, can use messages designed to
install \textit{Bloom filters}.
Once a Bloom filter is installed, the relaying node only sends matching object
information to the receiving node.

Each node in the network maintains a {\it mempool}, a local version of the
memory pool containing unconfirmed transactions. Nodes use the \textit{mempool}
message to request the contents of the receiving node's memory pool of
transactions.
In our experiments, we use the \textit{mempool} message along with Bloom
filters in order to efficiently check the reach of our messaging transactions.

\vspace{-15pt}

\section{System and Adversary Model}
\label{sec:model}

\begin{figure}[t]
\centering
\includegraphics[width=0.95\columnwidth]{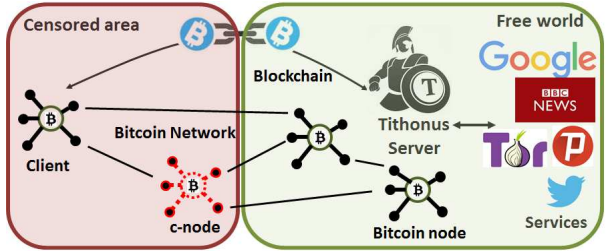}
\caption{Illustration of system and adversary model. The Tithonus server
leverages the Bitcoin network and blockchain, to provide censored clients with
access to news, and bootstrapping information for other censorship resistance
tools (source code, bridge IPs). The censor can control Bitcoin nodes and
filter detected suspicious transactions.}
\label{fig:model}
\vspace{-15pt}
\end{figure}

We consider a system that consists of several participants, see
Figure~\ref{fig:model} for an illustration. The {\it user} runs a Tithonus \textit{client} and a Bitcoin node, to retrieve information from a {\it target
destination}, e.g., content publisher. We assume that the user has control over
her computer and can install arbitrary software. The node is only initially needed in order to
access the Bitcoin network and blockchain, and retrieve the Tithonus client and
keying material. However, we recommend the use of a full Bitcoin
node even when the Tithonus client is not in use. Security reasons that we
describe in the following sections are the basis for this recommendation.
If storage and computational resources are a concern, the user
may choose to sacrifice provable security for convenience and only ``simulate''
the use of a Bitcoin full node. This option may be
susceptible to problems similar to the ones described by Houmansadr et al. in
\cite{houmansadr2013parrot}.

The {\it censor} is able to control (inspect, inject, suppress) the Internet
communications of users within an area, e.g., country. The censor can
fingerprint destination IPs, message contents and protocol semantics. We assume
that censored clients cannot use other censorship-resistant tools.  We assume
that the censor is not willing to block or significantly hinder cryptocurrency
use, and does not have the resources to modify or suppress access to any part
of the blockchain, i.e., it does not control a majority of the network
hashrate.

We assume that the target destination does not participate in censorship
evasion efforts. Instead, the {\it Tithonus service}, located outside of the
censored area, communicates with clients to assist them in connecting to target
destinations. We assume a single Tithonus service,
that may control multiple servers. We assume that the Tithonus service and its
servers are trusted to follow the protocol and not perform Sybil, DoS or packet
dropping attacks. We assume that the censor cannot identify the IPs of nodes
used by Tithonus, and cannot eclipse~\cite{HKZG15} the Bitcoin nodes controlled
by Tithonus.

Further, the external server can use anonymizers, e.g.  Tor, when
contacting exchange web-servers but never to perform Bitcoin transactions. This
policy avoids issues arising from using Bitcoin over
Tor~\cite{biryukov2015bitcoin}.

We assume that the censor is aware of Tithonus, and can register any number of
clients. In the following, we seek to protect only the communications of
Tithonus clients whose devices have not been corrupted (e.g., installed malware
on) by the censor.

\textbf{C-nodes}. We assume a censor who controls and deploys {\it
c-nodes}, Bitcoin nodes that can monitor events propagated through the network.
We introduce the {\it c-node filter} attack, where the censor allows censored
nodes to only connect to peers inside the censored area, but not to nodes
outside the censored area. C-nodes can connect to outside nodes, and their
purpose is to detect and hinder Tithonus transactions. However, we assume that
such disruptions are only directed towards interfering with the Tithonus system
(e.g., corrupt information communicated through Tithonus, degrade its
performance, including DoS attacks), and therefore actively avoid causing
disruptions to the Bitcoin network.

We further consider a powerful censor that attempts indirect user
identification by correlating Tithonus actions with transactions originated
within the censored region. For this, the censor needs to deploy
\textit{snooping c-nodes}, honest-but-curious nodes outside the censored
region. Otherwise, if the censor is not allowed to place nodes outside the
censored area, an incredibly simple countermeasure against such
\textit{snooping nodes} would be to avoid peering with nodes within the
censored region. Snooping c-nodes may try to peer with Tithonus nodes and
collect transaction timestamps or spending patterns for later analysis.

Thus, we consider an adversary who is both internal and
external. Specifically, the adversary is able to both (1) observe and control
the entire network within the censored area, and (2) deploy nodes on the
uncensored area, but not eclipse the Tithonus service.

\noindent
{\bf DoS and packet dropping attacks}.
We further consider adversaries that launch denial of service (DoS) attacks
against Tithonus, e.g., through excessive spurious/incomplete requests.
However, since we assume an adversary unwilling to disrupt the Bitcoin network,
we consider indiscriminate packet dropping attacks to be outside our threat
model as we assume an adversary unwilling to disrupt the Bitcoin network.

\vspace{-15pt}
\subsection{Solution Requirements}
\label{sec:requirements}

\begin{figure}[t]
\centering
\includegraphics[width=0.95\columnwidth]{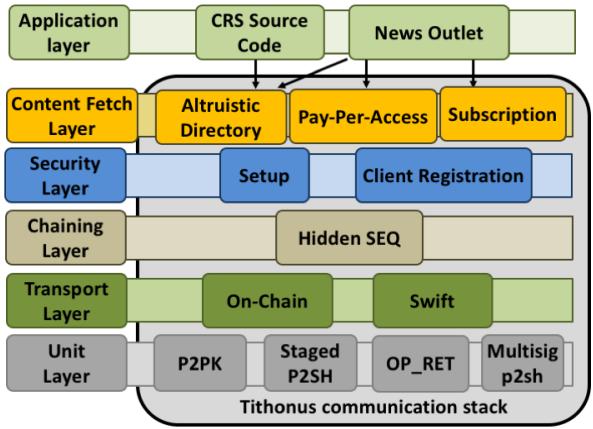}
\caption{Tithonus communication protocol stack. Each layer uses the layers
below to: communicate data units efficiently, transfer data of arbitrary size,
establish secure channels, request and retrieve content. The application layer
uses Tithonus to provide access to censored news, and other censorship
resistance tools.}
\label{fig:tithonus}
\vspace{-15pt}
\end{figure}

We distinguish two directions of communication, the {\it in-to-out}
communications are from censored clients (in) to the Tithonus server (out),
while {\it out-to-in} communication are on the reverse direction.

We seek to build a Bitcoin based censorship-resistant system that satisfies
several properties. {\bf Unobservability}: The censor is unable to detect
communications between the user and the publisher, even if it inspects the
packets sent and received by the user, their size, timing and destination.
{\bf Unblockability}: The censor is unable or unwilling to block communications
between the user and the publisher, even if it is able to identify such
communications. {\bf Availability}: The solution is resilient to DoS attacks.
{\bf Communication integrity}. The censor is unable to modify the
communications between the user and the target destination. {\bf Ease of
deployment}: The solution is easy to bootstrap and deploy, and does
not require altruistic participation. {\bf Performance}: The solution
minimizes costs and maximizes goodput.

\vspace{-15pt}
\section{Tithonus}

\begin{figure}[t]
\vspace{-15pt}
\renewcommand{\baselinestretch}{0.5}
\begin{minipage}{0.49\textwidth}
\begin{algorithm}[H] 
\begin{tabbing}
XXX\=X\=X\=X\=X\=X\= \kill

1.{\mbox{\bf{P2SH.scriptSig(data)}}}\\ 
2.\>{OP\_PUSHDATA2 data.getNext(520)}\\ 
3.\>{OP\_PUSHDATA2 data.getNext(520)}\\
4.\>{OP\_PUSHDATA2 data.getNext(520)}\\
5.\>{OP\_PUSHDATA1 \% push next 2 lines, 79 bytes}\\
6.\>{OP\_PUSH data.getNext(75)}\\
7.\>{OP\_2DROP OP\_2DROP OP\_TRUE}
\vspace{-55pt}
\end{tabbing}
\caption{{\it scriptSig} of p2sh input of the writing transaction in the
staged transaction writing method. Lines 4-6 are the {\it redeemScript}.} 
\label{alg:staged}
\end{algorithm}
\end{minipage}
\vspace{-15pt}
\end{figure}

\noindent
{\bf Overview}.
Most writing solutions surveyed in Section~\ref{sec:background:writing} have a
high per-byte cost of communication (see $\S$~\ref{sec:evaluation:block}), and
low goodput, as a transaction needs to wait around 10 minutes until it can be
mined into the blockchain. We introduce solutions that reduce the
time and per-byte cost of transmitting data through Bitcoin transactions. In
the following, we use {\it data unit}, to denote data that can be transmitted
in a single transaction, and {\it sender} and {\it receiver} to refer to the
communicating parties (censored client and Tithonus server).

We organize Tithonus into 5 layers, see Figure~\ref{fig:tithonus}. The unit
layer embeds data units into transactions, see
Section~\ref{sec:background:writing}. The transport layer optimizes transaction fees choices as to minimize the latency and cost per byte of content transmitted depending on the communication needs.
The chaining layer communicates messages of arbitrary size. The security layer
sets up the Tithonus root of trust, registers clients and establishes secure
communication channels with the server.  The content fetch layer enables
clients to securely retrieve desired content through Tithonus.  Tithonus
provides access to static and dynamic content, with free, pay-per-access and
subscription based payment options. As is the case for
layering architectures~\cite{CT1990}, each module within a layer is
independently combined with a module in a layer above it. For instance,
Tithonus can use the Multisig construct from the Unit Layer with the On-Chain
module from the Transport Layer.

\noindent
{\bf Design choices}.
To avoid detection and censorship, client initiated requests need to have
minimal size, and fit in a minimal number of transactions, of popular type. To
prevent DoS attacks, each client request needs to be backed by previously deposited user funds. 

\vspace{-5pt}

\subsection{Unit Layer}
\label{sec:tithonus:unit}

\noindent
{\bf Making sense of staged transactions}.
Staged transactions described in Section~\ref{sec:background:writing} have
never been properly documented, and older methods are likely no longer valid.
For instance, previous documentation mentions that one can write up to 10,000
bytes in an input script while only 1,650 are effectively
possible~\cite{scriptSigActualSize}. Further, Bitcoin often implements new
standard rules that modify the validity of new transactions. In its latest
implementation: (1) The writer can only push 520 bytes at a time on the Bitcoin
virtual machine stack. (2) The input script can only be 1,650 bytes in total,
thus one can push a bit over 3 chunks of 520 bytes, but the last push operation
needs to be the {\it redeemScript} in a p2sh redeeming transaction.  (3) The
redeemScript needs to empty the stack and push a true value before exiting.

Algorithm~\ref{alg:staged} presents the writing transaction's \textit{scriptSig} script of  the staged transaction writing method
($\S$~\ref{sec:background:writing}). The first three lines push on the stack
the first three 520 byte chunks of data.  Lines 5-7 are the {\it redeemScript},
that is first pushed on the stack (line 5), then pushes the last chunk of 75
remaining bytes of data (line 6) and then executes two double drop operations
that consume a total of 4 items from the stack. The final operation pushes a
TRUE value on the stack for the script to finish successfully (line 7).
This \textit{scriptSig} writes data units of up to 1,635 (i.e., 3 $\times$ 520
+ 75) \textit{effective} bytes.

\noindent
{\bf Multisig p2sh construct}.
The writing technique described above is easily detectable (transactions with
large input script sizes are uncommon), making it suitable for uncensorable,
out-to-in communications only. We propose now an inconspicuous, in-to-out
alternative that uses a ``MULTISIG'' pattern in the redeeming script to conceal
Tithonus transactions among normal m-out-of-n multi-signature p2sh
transactions. For Tithonus, we choose a 1-3 MULTISIG construct as it was traditionally the most common type of multi-signature transaction \cite{transactionTypes}. However, since the popularity of these constructs may change over time, Tithonus needs to be able to adapt to such changes. As a consequence, we describe techniques that are applicable to any MULTISIG construct but use the 1-3 MULTISIG as an example. In a 1-3 MULTISIG construct, the \textit{redeemScript} contains
placeholders for 3 public keys authorized to redeem the staging transaction. In
a normal transaction, the rest of the \textit{scriptSig} contains a valid
signature corresponding to any one of these public keys. In a Tithonus
transaction, the other two public key placeholders are instead filled with
arbitrary payload data. Compressed Bitcoin public keys are 33 bytes (including 1 \textit{prefix type}
byte).
However, these keys correspond to points of an elliptic curve, thus
are distinguishable from encrypted data.

\textbf{Embedding encrypted data on public keys.} To generate elliptic curve points
indistinguishable from random strings, we use ideas similar to the ones
described by Bernstein et al.~\cite{bernstein2013elligator}. Specifically, for
curve secp256k1, we propose the use of the left most significant 28 bytes to
encode our encrypted data $D$, with the exception of the ciphertext $2^{224}-1$
which has a negligible probability of occurrence.  
We
pick a random 4-byte string $R$, smaller than 0xFFFFFC2F. Thus, we get $x = D, R$, a
32-byte candidate for the elliptic curve point's $x$-coordinate that is smaller
than $p$, the prime used by sec256pk1.

We then calculate $w = x^3 + 7$ and check if it is a quadratic
residue (QR) in $\mathbb{F}_p$. We choose random $R$ values and repeat this
process until $w \in$ QR($\mathbb{F}_p$).  Since a random string has a $1/2$
chance of being a QR for secp256k1's prime field, the expected number of trials
is 2.
Once we have found a suitable $w$, we use the
corresponding $x = D, R$ value as the $x$-coordinate of the elliptic curve
public key. We use $x$ with a prefix of either 0x02 or 0x03 to signal
that this is the compressed representation of the public key.
Thus, this procedure embeds 28 bytes of encrypted data indistinguishable from
an elliptic curve public key. A proof sketch of this claim is provided in section \ref{sec:analysis}.

\vspace{-5pt}

\subsection{Transport Layer}
\label{sec:tithonus:transport}
\noindent
{\bf Swift transactions}.
The goal of the Bitcoin network is to distribute newly created transactions to
all the Bitcoin nodes across the world ($\S$~\ref{sec:background:network}).
Thus, a receiver who runs a node does not need to wait until the transaction is
mined into the blockchain, in order to access it. However, before propagating a
new transaction to its peers, each node verifies its validity, i.e., that (1) the
transaction fee equals or exceeds 1 satoshi per byte and (2) the value of each
output exceeds 3 times the transaction fee.

We leverage these observations to propose a low cost, low latency {\it swift
transaction} communication solution, that works with all of the communication
methods above and of $\S$~\ref{sec:background}. Swift transactions use minimal
transaction fee rates and values sent to non-data storing outputs (e.g., p2pkh
outputs, see $\S$~\ref{sec:background:writing}).  Swift transactions achieve
low latencies as nodes controlled by Tithonus clients and the server will
receive such transactions as they are propagated through the Bitcoin network,
well before they are mined into the blockchain.

\noindent
\textbf{On-Chain transactions}. The advantages of swift
transactions are fully exploited only when the user is online. For asynchronous
cases, on-chain transactions can increase the transaction fee to secure the
data in the blockchain as soon as possible. This increases the chances that the
user can access the information when they are online. On-chain transactions
thus trade costs for highly reliable communications.

\vspace{-5pt}

\subsection{Chaining Layer}
\label{sec:tithonus:chaining}

\vspace{-5pt}

The chaining layer sends messages of arbitrary sizes, by signaling the message
length, and the location and order of transactions storing individual data
units sent through the unit layer. Solutions like Catena~\cite{TD17} and
Blockstack~\cite{ANSF16}, are unsuitable to provide efficient censorship
resilient communications.

We propose a {\bf hidden sequencing} solution to address these problems.
The sender needs to embed into each data unit, the type of data (e.g., DATA, DIR, CERT, CREG, REQ) written by the layers above, see $\S$~\ref{sec:tithonus:security} and~\ref{sec:tithonus:fetch}).
The sender prefixes each of these data units with a 4 byte sequence number
$SEQ$, that specifies the order of the data unit within the content. In the
content's first data unit, the sender also includes a 4 byte $LEN$ value, which
specifies the number of data units in the content. If the content fetch layer
specifies a key, the sender uses it to encrypt each unit. The sender
writes each data unit into a swift transaction, e.g., in the p2sh input of a
staged transaction, or a p2pkh output (see
$\S$~\ref{sec:tithonus:unit} and~\ref{sec:background:writing}), then injects
all transactions in the Bitcoin network, in parallel. The sender can process
multiple contents simultaneously, using a locally stored, atomically accessed
$SEQ$ variable.

The receiver needs the 32 byte id of the transaction that stores the first data
unit, and optionally, a decryption key. He recovers the $SEQ$ and
$LEN$ values from the first unit, then accesses (and optionally decrypts) all
p2sh inputs and p2pkh outputs of transactions in the mempool and nearby blocks
in the blockchain, until he recovers all $LEN-1$ subsequent $SEQ$ numbered data
units (see $\S$~\ref{sec:tithonus:security} and~\ref{sec:tithonus:fetch} for
more details).

\vspace{-5pt}

\subsection{Security Layer}
\label{sec:tithonus:security}

\vspace{-5pt}

The security layer enables clients to establish trust, and to register and
establish secure communication channels with the Tithonus service.

While suspicious out-to-in transactions cannot be blocked, as they will
eventually be persisted in the blockchain ($\S$~\ref{sec:evaluation}), high
numbers of client posted transactions are intrinsically suspicious and can be
discovered and blocked by a c-node filtering censor ($\S$~\ref{sec:model}).
Thus, in the following, we seek to minimize the number of transactions posted
by clients.

\noindent
{\bf Setup}.
The Tithonus service creates a self signed public key certificate and uses the
chaining layer to publish the certificate on the Bitcoin blockchain (data
units of type CERT, see $\S$~\ref{sec:tithonus:chaining} and
Figure~\ref{fig:altruistic:dir}). We call this the {\it root certificate}.
While an adversary may also write a certificate impersonating Tithonus, the
root Tithonus certificate is the oldest one.
Let $pk_T$ and $sk_T$ be the public and private keys of the Tithonus service.
The certificate specifies the key agreement, key derivation, hash, symmetric encryption and message authentication code functions used (e.g., ANSI X9.63~\cite{ANSI.X9.63}).  
The certificate also includes a random tag value $Ttag$ and a standard fee rate, i.e., price in satoshis per written byte, for responding to client requests.

Tithonus publishes new certificates periodically, see
Figure~\ref{fig:altruistic:dir} for an illustration. A new certificate is
signed with the private key of the previous certificate. The client is
responsible for finding all the certificates, starting from the oldest to the
newest one, and verifying the chain of trust. For simplicity, the Tithonus
certificate also includes a $SEQ$ number, set to 0 for the root.

Further, we use the chaining layer to write the Tithonus client on the
blockchain.  This allows censored users to download the client with as little
information as the id of the first transaction storing it.

\noindent
{\bf Client registration}.
Upon startup, a client $C$ inside the censored area follows ECIES~\cite{S01} to
establish a session key with the Tithonus server through an encrypted client
registration ($CREG$) message. Specifically, the client generates a private key
$sk_C \in_R \mathbb{Z}_n^*$ and a public key $pk_C = sk_C G$, where $G$ and $n$ are as specified in secp256k1. 
It then
generates a Rijndael (28B block mode) encryption key $K_1$ and a hashing key $K_2$, using the KDF
specified in the Tithonus certificate with input $S_x$, where $(S_x,S_y) = sk_C
pk_T$ is a point on the elliptic curve (shared secret) and $pk_T$ is the public
key of the Tithonus server. $C$ then generates a random session tag identifier
$R_{CT}$ to identify subsequent communications with the Tithonus server, and a
``payment'' Bitcoin address with public and private keys $pk_{fee}, sk_{fee}$.
The client sends to the Tithonus server, the message:
\[
pk_C, E_{K_1}(Ttag, CREG, R_{CT}, sk_{fee}, padding)
\]
where $CREG$ is a 1 byte long header type that differentiates client
registration messages from other Tithonus communications. If the client needs
to re-register (e.g., this procedure fails), the client needs to generate new
keys to avoid observable duplicates. The AES encryption uses a function of
$pk_C$ as initialization vector. When using ECIES, without padding, this
message is 98B long, assuming a $Ttag$ and $R_{CT}$ size of 16B each,
$sk_{fee}$ of 32B, $pk_C$ of 33B compressed form, and $CREG$ of 1B.

The client camouflages this message into 2 staged multisig p2sh transactions
($\S$~\ref{sec:tithonus:unit}), where each staging transaction has 2 outputs, of
type ``p2pkh, p2sh''.  Figure \ref{fig:output:distrib} (Section \ref{sec:background:writing})  shows that such
transactions are the second most popular, thus do not stand out. Further, a
client registers infrequently. We write the above message into the redeeming
script sections of the p2sh inputs.  Each of these inputs has a net capacity for storing 2 * 28 bytes ($\S$~\ref{sec:tithonus:unit}, MULTISIG script).
However, the $pk_C$ (33 bytes) element is stored in one of the MULTISIG
addresses, leaving 28 bytes available in one of the p2sh inputs. The second
p2sh provides another 56 bytes for a total of 84 bytes available for the second
part of the $CREG$ message. Thus, the random $padding$ needs to be 19B (84 -
65) long.

The staging transactions have two p2pkh outputs. The first output
is used to fund the second staging transaction thus links the 2 transactions.
The second output contains the fee (in satoshi), which needs to cover the
server's expenses to write the reply in the blockchain.

To identify client registration messages, the Tithonus server processes all
pairs of p2pkh;p2sh transactions. Let $T$ be a pair of writing
transactions with p2sh inputs $I_1$ and $I_2$, and outputs $O_1$ and $O_2$. The
server concatenates the 4 unused public keys placeholders from $I_1$ and $I_2$
(132B), reads the first 33B as candidate $pk_C$, strips the prefix bytes from
the remaining 3 public key placeholders, generates candidate point $(S_x, S_y)
= pk_C sk_T$, then uses it as input for the KDF function to construct candidate
keys $K_1$ and $K_2$. The server decrypts the remaining 84B (99B - 3 prefix
bytes - 12 random pad bytes) using $K_1$. If the result does not start with ($Ttag, CREG$), the
server drops the transaction, as it is either not a $CREG$ message or is
corrupted. Otherwise, the server recovers from the decrypted message the
session tag $R_{CT}$ and the private key $sk_{fee}$.

The server uses $sk_{fee}$ to compute the public key $pk_{fee}$.  It then
compares $H(pk_{fee})$ against the public key hash value stored in the p2sh
output $O_2$ of transaction pair $T$. If the verification fails, the server
drops the transaction. Otherwise, the server uses $sk_{fee}$ to redeem the fee
from $O_2$, and creates a record for client $C$:
\[
R[C] = [pk_C, K_1, K_2, R_{CT}, H_{K_2}(R_{CT}, c), c=1, credit] 
\]
$R[C]$ contains the session and hashing keys, the fresh session tag, the
message count denoting how many messages it has exchanged with the client, and
the balance it has redeemed from the user's payment. 
The
server updates $R[C]$ each time it processes a message to/from client $C$:
it increments $c$, decrements $credit$ and updates the tag $H_{K_2}(R_{CT}, c)$.

Tithonus allows clients to register multiple times, creating additional records
for new $pk_{fee}$ keys. This prevents observability, by reducing the amount of
repeated deposits to the same payment account. It also prevents Tithonus from
learning detailed, client request profiles. 

\vspace{-15pt}

\subsection{Content Fetch Layer}
\label{sec:tithonus:fetch}

\vspace{-5pt}

Clients use the content fetch layer to request and fetch censored content. We
introduce solutions to fetch static and dynamic content, with free,
pay-per-access and subscription based payments.

\begin{figure}
\centering
\includegraphics[width=0.95\columnwidth]{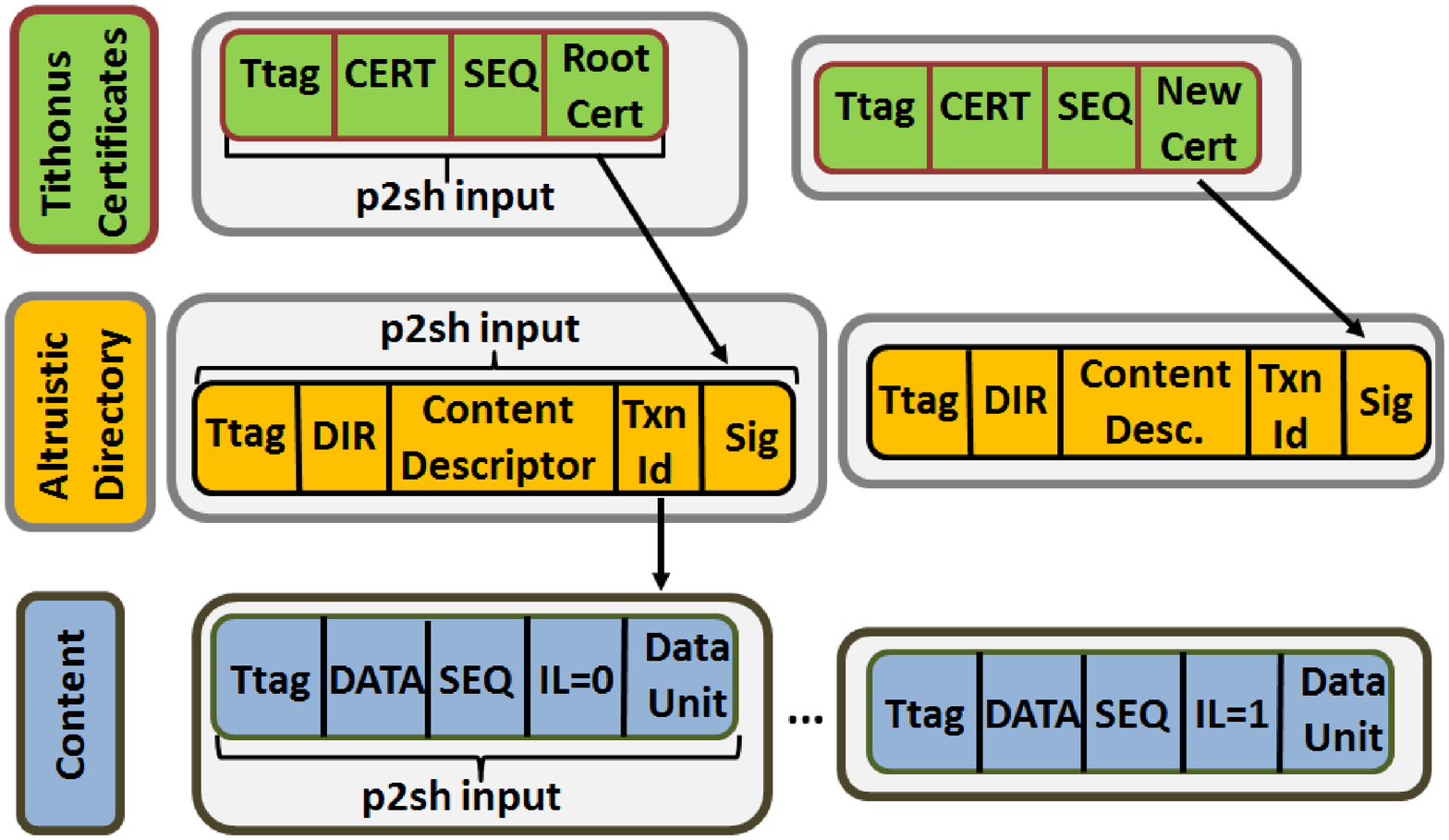}
\caption{Altruistic directory organization. Client reads all certificate (CERT)
transactions and verifies the chain of trust. Client then reads all directory
entry (DIR) transactions and verifies that they are signed with the certificate
that was valid when they were issued. DIR entries point to the first DATA
transaction storing the content, which is sequenced by SEQ numbers.}
\label{fig:altruistic:dir}
\vspace{-15pt}
\end{figure}

\noindent
{\bf Altruistic directory: free, static content}.
The altruistic directory solution, illustrated in
Figure~\ref{fig:altruistic:dir}, allows clients to access censored content that
is considered by Tithonus to be of public interest, e.g., the source code of
other censorship-resistant systems (CRS), news articles, for free.  Content is
written in cleartext on the blockchain. Its authenticity is ensured through a
signature with the most recent Tithonus private key. The client can fetch
content listed in the altruistic directory without needing to register with the
server (see $\S$~\ref{sec:tithonus:security}).

Specifically, for each new piece of content that Tithonus decides to distribute
for free, the server uses a two-step process. First, use the chaining layer to
write the compressed content into the blockchain (data units of type DATA, see
bottom part of Figure~\ref{fig:altruistic:dir}). As consumers for this content can not be expected to be online at the time of publishing, swift transactions are inadequate and the On-Chain transport module is preferable. 
Second, the server generates a new {\it directory entry}, that
contains (1) a description of the content, (2) the id of the leading
transaction in the blockchain, that stores the first data unit of the
content, and (3) the Tithonus signature over the previous fields,
with the most recent Tithonus private key (middle layer in
Figure~\ref{fig:altruistic:dir}). It then uses the chaining layer to write
this new directory entry into the blockchain, as data units of type DIR.

The client needs to retrieve and reconstruct the entire directory. To reduce
the need of clients to parse each Bitcoin transaction looking for DIR entries,
the server can write DIR entries into transactions using a specific input
Bitcoin, included in the Tithonus certificate. The client then needs to only
retrieve and parse a subset of the transactions to retrieve DIR entries. The
client verifies the authenticity of DIR entries: the signature in each entry
was generated using the public key stored in the Tithonus certificate that was
valid when the DIR's block was mined into the blockchain.

If the client wants to fetch a specific content listed in a DIR entry, it
retrieves the id of the first transaction that stores the content's DATA units,
from its DIR entry, then uses the chaining layer to read the content from the
blockchain ($\S$~\ref{sec:tithonus:chaining}). The client monitors the
blockchain to identify newly added directory entries.

\noindent
{\bf On-demand, pay-per-access}.
We propose a content fetching approach where the client can unobservably
request new content, then fetch it with confidentiality and integrity
assurances, while the server can charge for writing new content, and cache
popular content, thus reduce the cost and observability of subsequent requests.
A client requests content through a $REQ$-type message:
\[
H_{K_2}(R_{CT},c++), E_{K_1}(Ttag, REQ, SEL, URI),
\]
where $H_{K_2}(R_{CT},c)$ is used to denote a fresh session tag for client $C$,
$c$ being $C$'s message counter. $c++$ signifies that the counter is
incremented by both the client and the server after processing this message.
$REQ$ is a 1B message type (request). $SEL$ is an 8B selector tuple (\textit{offset}, \textit{length}) that requests a specific offset (4B) and length (4B) of the content. $URI$ is the \textit{null-padded} pointer to the content, which could be a known site (e.g., \url{www.bbc.com}) or a tinyurl returned by the server for a previous request (e.g., Google search results). The client uses 2 staged
multisig p2sh transactions as in the client registration, where the staging
transaction has 2 outputs (p2sh; p2pkh). The client camouflages the above
message into the p2sh inputs (112B). The $H_{K_2}(R_{CT},c++)$ is embedded in
the first 28B of one of the p2sh inputs with 8 bytes of random padding. Thus, since
$Ttag$, $REQ$ and $SEL$ require a total of 25B, the null-padded $URI$ can be
up to 59 bytes long (84 - 25). Just as in the case of the client registration,
the first p2pkh output is used as a link to the second staging transaction
while the second p2pkh is not used. The reason for choosing this
transaction type is that
it is the second most popular, accounting for $\approx$ 19\% of all
transactions when counting p2sh;p2pkh and p2pkh;p2sh inputs and outputs, see
Figure~\ref{fig:output:distrib}.

To identify $REQ$ messages, the server processes each pair of p2sh; p2pkh output transactions linked by the first p2pkh output. For each candidate pair, the server extracts the first 28B from the first p2sh input and concatenates the remaining 28 byte chunks of the remaining p2sh inputs. It then looks up the first 20B ($H_{K_2}(R_{CT},c)$) of the result, among the session tags of all hosted clients
($\S$~\ref{sec:tithonus:security}).  When it finds a match, the server
increments the client's $c$ count, recovers $C$'s $K_1$ key and decrypts the
remaining 84B ($E_{K_1}(Ttag, REQ, SEL, URI)$).  The server checks the
user's record balance, 
then fetches the content at the
specified $URI$, determines the cost $fee$ required to send it through Tithonus
(e.g., based on content size, whether it is cached or not, predicted
popularity, see below) and sends the following reply:
\[
Ttag,H_{K_2}(R_{CT},c++), E_{K_1}(fee, RESP),rev(Ttag)
\]
The response is embedded in a chain of p2sh staged transactions, each storing
1,635 bytes ($\S$~\ref{sec:tithonus:unit}). The answer's length is dependent on
the available user's balance. If this balance is insufficient, the server
answers only the number of bytes covered by the balance starting at the
specified offset. Otherwise, the request is ignored. The client needs to process
all transactions chains marshalled by a
$Ttag$ and its reverse ($rev(Ttag)$), both in the mempool and newly posted in
the blockchain, in order to identify the server reply. Once a reply is
identified by the $H_{K_2}(R_{CT},c++)$ tag, the client decrypts the rest of the
chain and retrieves the requested content.

\noindent
{\bf Subscription based access}.
Tithonus can convert the altruistic directory and cached content fetch
solutions into subscription based solutions, to provide access to dynamic
content, e.g, news services, RSS feeds. The client subscribes interest in
published content (e.g., through a $URI$ as above), and transfers funds to the
Tithonus server. The server periodically accesses the content, publishes
updates using the chaining layer, then updates the client's balance. The server
may charge less per update if multiple clients subscribed to this content. The
subscription can be update or time based, i.e., the client pays per update or
time unit. For this, the service encrypts content with a key that it
distributes (encrypted) to subscribed clients. The key can change after each
update, or periodically (e.g., once per day).

\section{Analysis}
\label{sec:analysis}

\vspace{-5pt}

In this section we analyze the ability of Tithonus to satisfy the
requirements outlined in Section~\ref{sec:requirements}.

\noindent
{\bf Unobservability.}
In the following we analyze the ability of Tithonus to protect the user's
unobservable access to out-to-in communication and to ensure the
indistinguishability of in-to-out requests.

\noindent
{\bf Access to out-to-in communications}.
User nodes need to exhibit full node functionality thus can access
free content published using the altruistic directory method, in an
unobservable manner: Full Bitcoin nodes can only function if they fetch the
entire blockchain, thus clients can just access the blockchain to read and
verify the Tithonus public key certificate chain and the altruistic directory,
then recover the desired content, all written in the blockchain.
Thus, the Bitcoin ecosystem separates clients from the Tithonus service, providing
potential for anonymity.  Client nodes download the entire blockchain. Thus,
the altruistic directory is equivalent to the trivial solution to the PIR
problem, and requires no direct client contact with the Tithonus service. In
the on-demand, pay-per-access solution, clients identify only through a random
public key. Clients can register multiple public keys (i.e.,
pseudo-identities), to prevent the server from building and de-anonymizing
profiles. 

For the staged transactions issued for the client registration, cached content
fetch and subscription based access method, we use the proposed patterns (e.g.
p2pkh, p2sh inputs and outputs) only as templates: Tithonus camouflages client
issued protocol messages in popular transaction types, whose distributions
change overtime. While, currently, such transactions account for 20\% of
Bitcoin transactions, see Section~\ref{sec:evaluation}, Tithonus adds inputs or
outputs depending on the evolution of these changes but preserves the use of
the underlying barebones transaction pattern.

Tithonus restricts the amount of content that a client can request per time
unit (e.g., per day) to a value consistent with that of regular Bitcoin users. In addition,
Payments issued by clients cannot be traced to the Tithonus service if at least
one of the exchanges employed by the server does not collude with the censor. 

\noindent
{\bf Indistinguishability of in-to-out requests}.
This communication type makes use of multisig p2sh transactions embeddings
(Section ~\ref{sec:tithonus:unit}).
Since this embedding directly uses the encryption output of a Rjindael cipher, 
its correctness follows from the correctness of
the Rjindael algorithm~\cite{duan2005functional}.
We include a proof sketch that the Tithonus multisig p2sh constructs are
indistinguishable from regular multisig p2sh transactions, in
Appendix~\ref{appendix:proof}.

\noindent
{\bf Unblockability}.
The system's resilience to censorship is based on the unwillingness of the
censor to affect the Bitcoin ecosystem's normal functioning, which stems from
reasons ranging from economic to technical. We detail these reasons in
$\S$~\ref{sec:discussion}. 

\noindent
{\bf Availability.}
The Tithonus server does not expose a traditional communication
end-point, thus its service is not vulnerable to traditional DoS attacks based on
excessive spurious/incomplete requests. All Tithonus requests need to be paid
upfront and thus any increase on the number of requests is met with more
resources afforded by the corresponding request fees.
These fees can also include operation, maintenance and profit fees for the
Tithonus service, converting a DoS attack into a wealth transfer from the
attacker to the Tithonus infrastructure. A censor that floods the Bitcoin
network with Tithonus transactions to exhaust its resources, will further lead
to congestion and transaction fee rate spikes, thus disrupt the entire Bitcoin
ecosystem, including e.g., e-commerce merchants and their customers.

\noindent
{\bf Communication integrity}.
The client authenticates the Tithonus server through its ability to decrypt
messages encrypted with the public key advertised in the Tithonus certificate.
The client also verifies that specific messages (e.g., DIR entries) are signed
with the private key of the Tithonus server. The server verifies that the
client has provided the funds required to send back the replies through the
Bitcoin network or blockchain. Both the client and the server use special
fields in messages exchanged, to verify their integrity and authenticity.
Further, the chaining layer preserves the order of the data units, and the use
of erasure codes can provide resilience to data filtering.

\noindent
{\bf Ease of deployment}.
Tithonus users only need to know the 32 bytes of the transaction id that stores
the first data unit of the Tithonus client source code, in order to first fetch
the source code, then compile and run it. Alternatively, a small Tithonus
client {\it bootstrapper} can perform these operations.

\noindent
{\bf Performance.} Tithonus uses minimal values for transaction
fess and adapts them according to the communication need (swift or on-chain
transactions). In addition, the use of staged transactions maximizes the
payload data output under the current Bitcoin transaction rules.
Next, we evaluate the performance of Tithonus.

\vspace{-15pt}

\section{Evaluation}
\label{sec:evaluation}

\noindent
{\bf Ethical considerations}.
In our experiments, we did not interact with humans. We have only collected the
country of location of nodes, and times when they were online.
In the following we discuss the potential burden placed by Tithonus on Bitcoin
miners and clients.

\noindent
{\bf Burden on miners}.
Given the high costs of the mining hardware, electricity and overall
maintenance, the main motivation for miner participation is financial.
Tithonus does not add to these costs. For out-to-in communication, Tithonus
transactions even add financial incentives to miners. Other blockchain-writing
services, e.g., Catena, Apertus, place a similar burden on miners and clients.
Thus, Tithonus does not expose them to any additional risk not already accepted
by them. This is an inherent side effect of collateral damage-based solutions.
In fact, mining activity has increased
exponentially~\cite{bitcoinhashrategrowth} despite the continued implementation
of such un-anticipated uses of the blockchain.

\noindent
{\bf Burden on clients}.
Users who run light clients, that do not fully participate on the
gossip protocol, are unburdened by Tithonus transactions.  Only clients that
run ``full nodes'' are impacted by Tithonus. While their motivation to do so
is transitively financial (i.e., to support miners who make a profit from
mining them), we admit that Tithonus may place an unwanted burden on purely
altruistic full nodes. In this respect, Tithonus is similar to other collateral
damage based CRS systems that e.g., use domain fronting or CDN caches.

\vspace{-5pt}

\subsection{Tithonus Certificate}
\label{sec:evaluation:block}

\vspace{-5pt}

We have written the first Tithonus certificate (714 bytes in compressed form)
using a staged transaction ($\S$~\ref{sec:tithonus:unit}). The resulting
transaction is available at \url{https://tinyurl.com/y8u4avu6}. The cost was
1000 satoshis (1.117 sat/byte fee rate, i.e. almost min rate) to write an 895
bytes transaction containing the certificate. The p2sh input includes a small
accompanying script that extracts the zip file from the raw transaction:
\[
\tt{echo\ rawTxn\ |\ dd\ skip=47\ bs=2\ count=714\ |}
\]
\[
\tt{sed\ 's/4cc0//'\ |\ xxd\ -r\ -p\ >\ tithonus\_cert.zip}
\]
The transaction made it to the blockchain in $\approx$ 7 mins.

\vspace{-15pt}

\subsection{Swift Transactions}
\label{sec:evaluation:swift}

\vspace{-5pt}

We implemented components of the Tithonus infrastructure using 790 Python loc.
We have prepared 8 p2pkh transactions with one input and two outputs.  We have
issued 4 types of transactions, each assigned a transaction fee of
1, 2, 4, and 8 times the minimum transaction rate $fee$ of 1 satoshi/byte. 

We downloaded a list of 924 Bitcoin nodes' IP addresses from earn.com from the
35 countries with least freedom of press according to reporters without borders
(rsf.org). Only 12 of these countries had Bitcoin nodes that
accepted incoming connections (``server nodes''). We use these nodes' relay
times to estimate the time for a Tithonus message to spread across the network.
However, Tithonus does not require server nodes for operation. Tithonus is
usable as long as the censor allows at least ``client'' nodes that peer to
other ``server nodes'' (outside or inside the censored region). We initiated a
connection with these server nodes inside censor areas. We were able to maintain connections to 530 nodes
over the entire duration of the experiment, i.e., 24 hours, by answering
\textit{ping} messages but ignoring other commands.

\begin{figure}
\centering
\includegraphics[width=0.95\columnwidth]{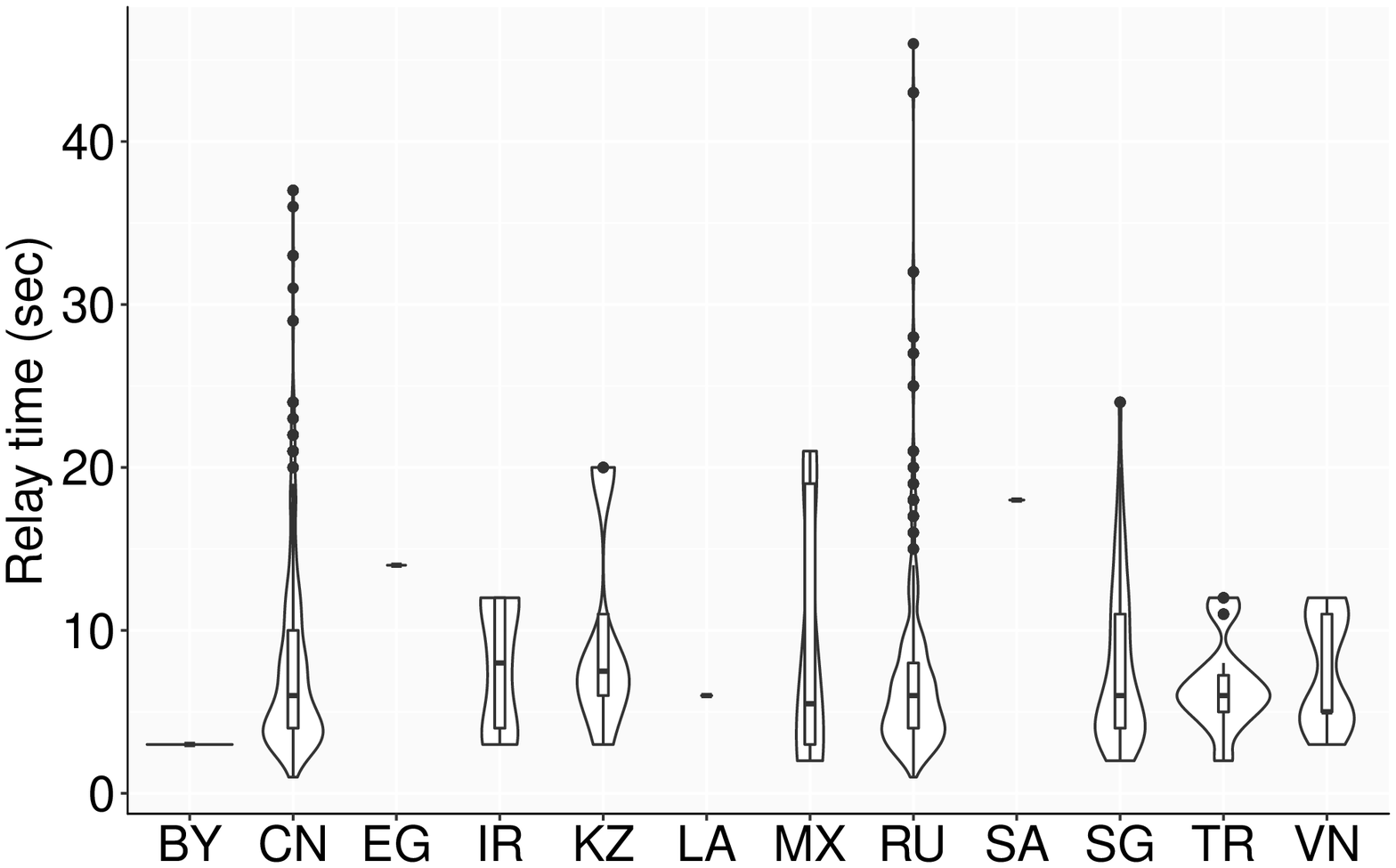}
\caption{Violins show the probability density of relay times (for all 1-8 fee
txns) over the nodes in each of 12 censored countries.}
\label{fig:txn:countries:all}
\vspace{-15pt}
\end{figure}

We set up a {\it sentinel node} to connect to all these 530 nodes, added Bloom
filters encoding our transactions, into all these nodes, and waited to receive
our transactions with a voluntary \textit{inv} message. We used
\textit{mempool} messages to retrieve the contents of non-relaying nodes'
mempool and verify they never received them.

We have set up a default installation of the Bitcoin reference client and
allowed it to connect to 8 peers outside the censored countries. We injected
our transactions from this node. We used the interval between the time when the
sentinel node received a transaction, and the time when the transaction was
injected into the network as an upper bound on the time our transaction took to
reach its destination.

526 nodes relayed at least one of our transactions (99.24\%), and
509 nodes relayed all 4 transactions.
Half of the nodes received our transactions in less than 5s and 90\% of the
nodes received them in under 20s.  Figure~\ref{fig:txn:countries:all} shows the
distributions of relay times over the nodes in each of 12 censored countries.
Most nodes in countries like China and Russia (who have the most nodes) relay
the transactions in under 5s, but have a few nodes who take longer than 30s
(but under 50s). All the nodes in the other countries relay our transactions in
under 25s.

\begin{figure}
\centering
\includegraphics[width=0.95\columnwidth]{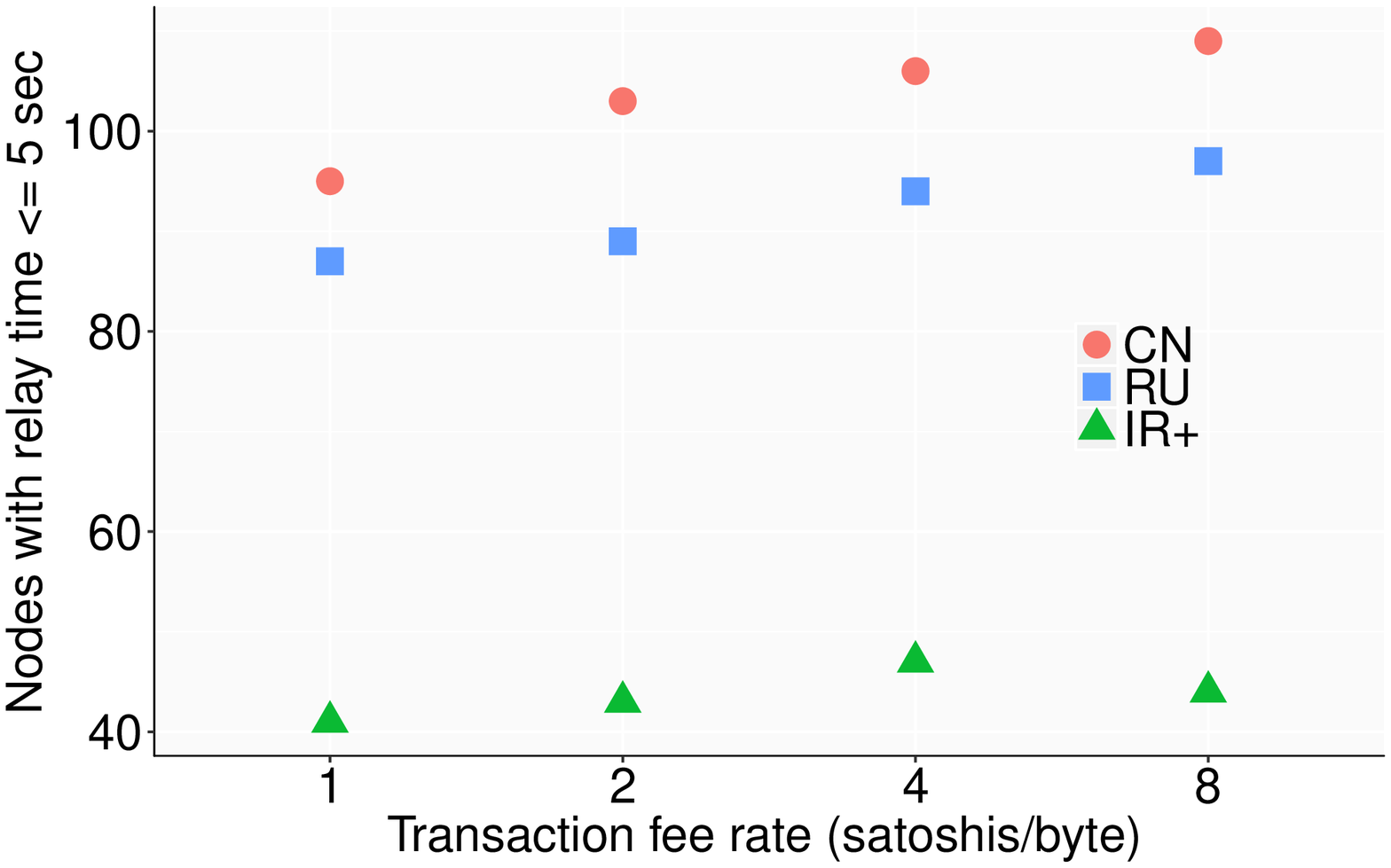}
\vspace{-5pt}
\caption{Number of nodes in China, Russia and the other 10 censored countries
together, that relay transactions in 5 seconds or less. Even at the lowest fee,
41.8\% of the Chinese nodes and 45\% of the Russian nodes relay the
transaction.}
\label{fig:txn:nodes:all}
\vspace{-15pt}
\end{figure}

A censored client will receive a transaction if at least one of its peers
receives it. To understand the ability of a censored node to receive low fee
transactions, we plot the number of nodes per country that relay each of the
1-8 fee transactions. Figure~\ref{fig:txn:nodes:all} plots these numbers for
CN, RU and the other 10 censored countries together, that relay our
transactions in 5s or less. We see a linear increase in the number of CN and RU
nodes who relay transactions, as a function of the transaction fee. 

\noindent
\textbf{Swift transaction's reliability and speed}. Even at the lowest fee, 41.8\% nodes in China and 45\% in
Russia relay the transaction in under 5 secs. Thus, if the Tithonus client
connects to 8 random peers in any of these countries, it will receive even the
lowest fee transactions with high probability in under 5 secs (e.g., 98.68\% = $1 - (1
- 41.8\%)^8$ in China).

\begin{figure}
\centering
\includegraphics[width=0.95\columnwidth]{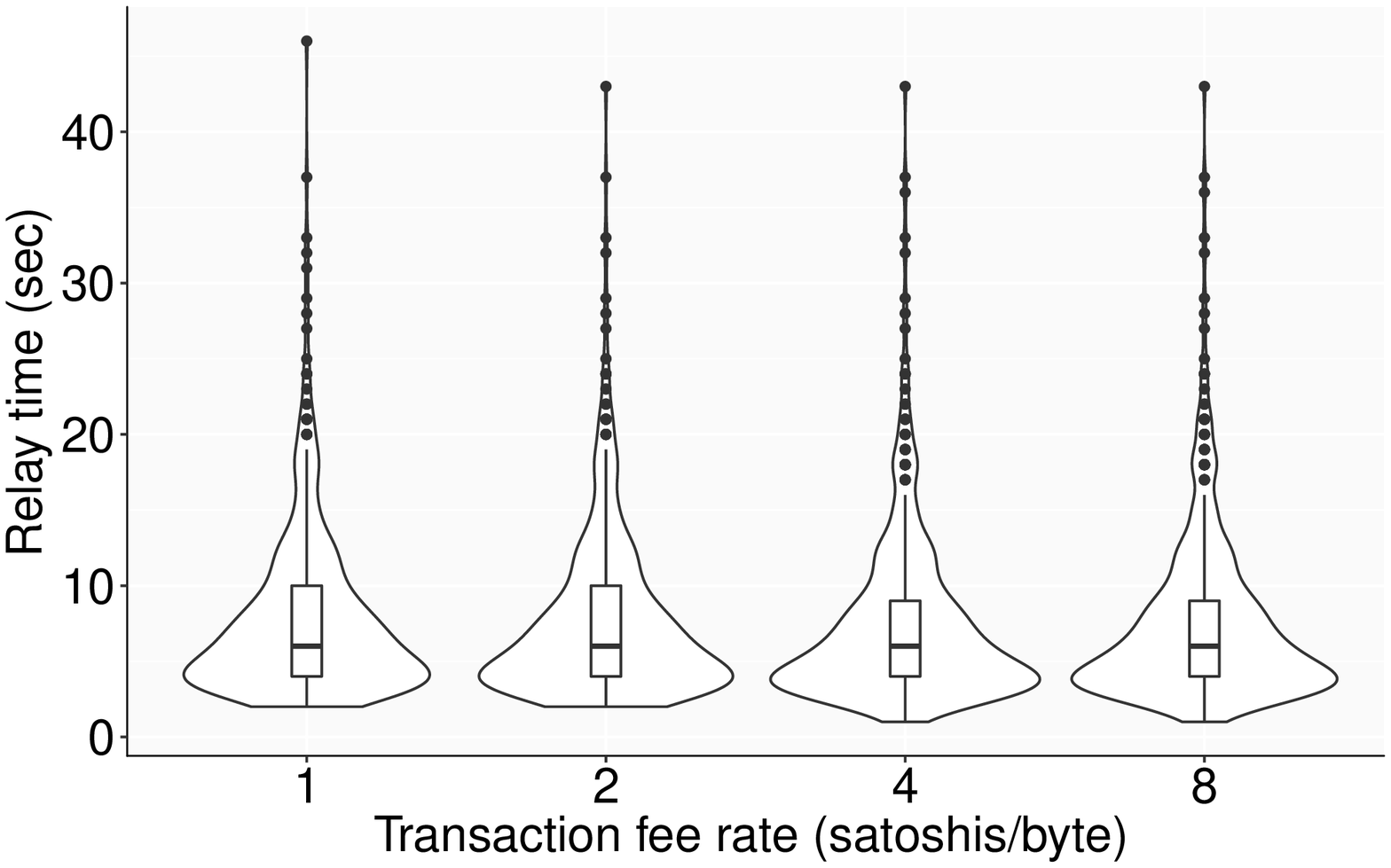}
\caption{Violins show the probability density of the relay time as a function
of the transaction fee rate, over all the nodes in the 12 censored countries. We
observe a decrease in relay times with the transaction fee. The median
is 6s over all transaction fees.
}
\label{fig:txn:fee:all}
\vspace{-15pt}
\end{figure}

Further, we have evaluated the ability of an increase in the transaction fee to
reduce the relay time of a message. Figure~\ref{fig:txn:fee:all} shows the
probability density of the relay times over all the nodes in the 12 censored
countries, for each of the 4 transaction fee rate types. The median relay time
remains constant at 6s when the transaction fee rate increases from 1 to 8
satoshis per byte. This suggests no advantage in reducing relay times by
increasing transaction fee rates in this interval.

\vspace{-15pt}

\subsection{The Price of Free Speech}
\label{sec:evaluation:comparison}

\vspace{-5pt}

\noindent
{\bf The cost of out-to-in communications}.
We have used Tithonus to send a 13,804B file (Tithonus logo) to nodes in the 12
censored countries. The file fit into 9 staged transactions, using a single
p2sh input per transaction, and the minimum transaction fee rate (1 sat/B).

\begin{figure}
\centering
\includegraphics[width=0.85\columnwidth]{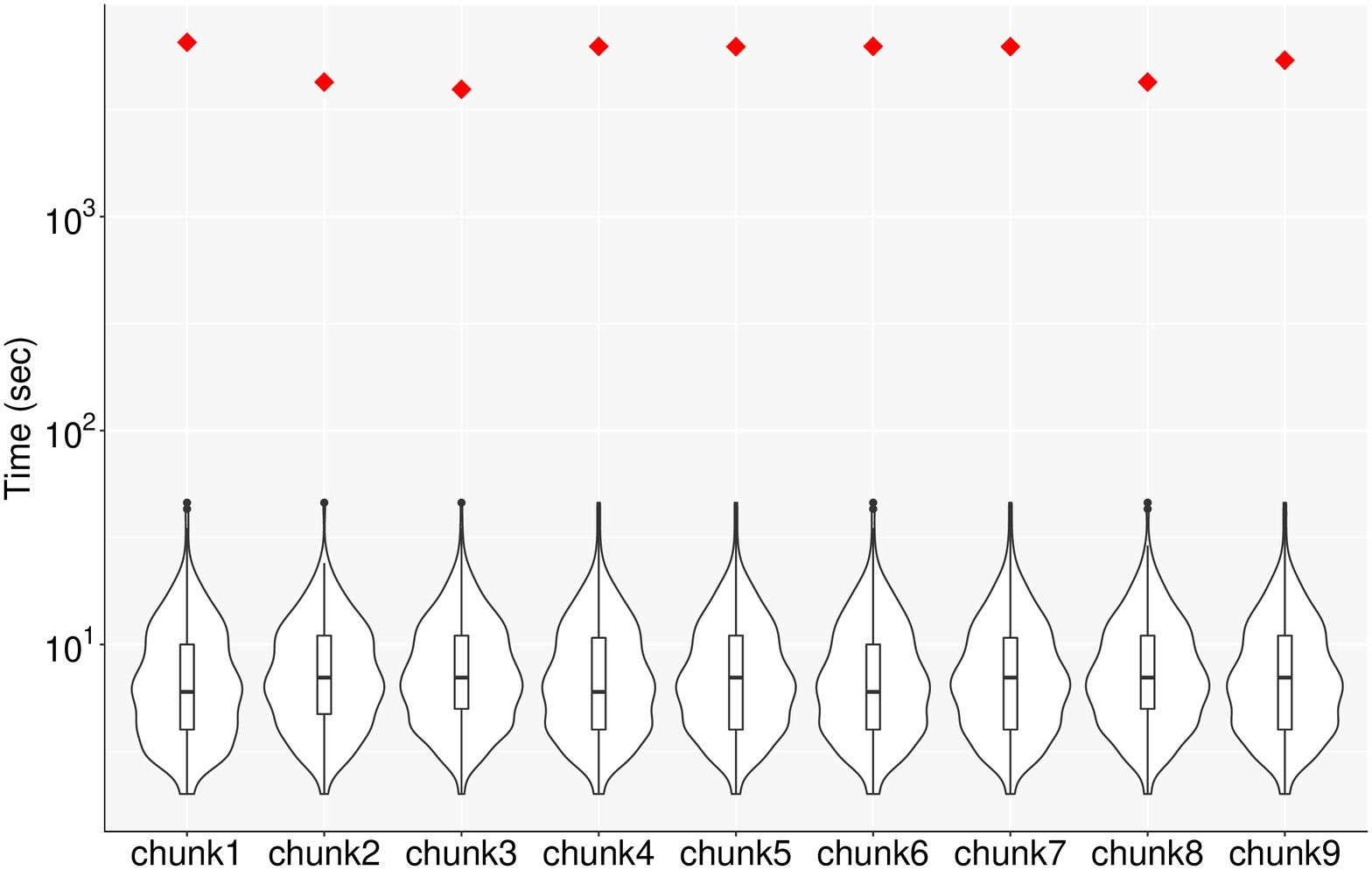}
\caption{Distribution of relay times by nodes in 12 most censored countries,
and blockchain arrival times for 9 transactions carrying the Tithonus logo.
Swift transactions are relayed by responding nodes in less than 40
seconds while all transactions eventually make it to the blockchain in less
than 1.8 hours.}
\label{writing:times:distribution}
\vspace{-15pt}
\end{figure}

We used the same setup as in the previous experiment and obtained similar results in terms of reliability. Figure~\ref{writing:times:distribution} shows the probability distributions of
the relay times of each of the 9 transactions, along with the time it took each
transaction to be mined into the blockchain.  All 9 transactions were
permanently recorded in the blockchain, in at most 1.8 hours. This result
suggests that filtering swift transactions is a useless censorship strategy:
censorship only delays
their delivery. In addition, selective filtering may harm pools and miners who do not mine these transactions~\cite{decker2013information}.

Further,
the total cost of writing the 9 transactions on the blockchain was 14,722
satoshi, which includes 8 transactions of size 1,656 bytes and 1 transaction of
size 1,474 bytes. 
At current prices (1 BTC $\approx$ \$\usd{\btcPrice}), the Tithonus cost of
sending the 14KB file was \$\satToUSD{14722}\result.

\noindent
{\bf Tithonus cost for an average web user}.
We consider now a scenario where the user accesses a news article of an average
of 1,200 words~\cite{Article.Words}. Assuming an average of 6 characters per
word \cite{wordLen}, 
the total
size of 7.2KB would require Tithonus to send 7.2 * 1.07 = 7.7KB, i.e.,
including transaction overheads. Therefore, the price to write this to the
blockchain is 7,704 satoshis or \$\satToUSD{7704}\result~(at the current rate of \$\usd{\btcPrice}  
per BTC).

\noindent
{\bf The cost of in-to-out communications}.
The client registration process requires the creation of two \textit{multisig
	transactions} for the registration process. Each subsequent resource request
message requires 2 additional \textit{multisig transactions}
($\S$~\ref{sec:tithonus:security}). A p2sh multi-signature transaction with 2
outputs (p2sh, p2pkh) has a size of 395 bytes. We leverage the estimates from
\cite{bitcoinfees} 
in order to determine a fee per byte that would allow our transactions
to blend in with other p2sh-multisig transactions. The median fee per byte over
the last 24 hours at the time of writing is \btcFee~ 
satoshi/B. Thus, the user needs
to pay a total of \ptwoshPrice{790}\usd{\result}  
satoshi ($\approx$ \$\satToUSD{6320}\result~
at the current conversion
rate) to register, and 
for each subsequent
content request message.

\begin{table}[t]
\centering
\small
\caption{\bf{Tithonus vs. VPN in-to-out cost and latency from
China, with separate conversation and communication setup phases.}}
\vspace{-5pt}
\resizebox{0.49\textwidth}{!}{
\textsf{
\begin{tabular}{l c c c c c c}
\toprule
Communication & Setup & Request & Communication & Download & Download & Download  \\
Service & Cost & Cost & Setup Time& Req. Time & Resp. Time &Total Time\\
\midrule
Pay-per-Access & \$\satToUSD{6320}\result & \$\satToUSD{6320}\result & 10 min. & 3-5s & 3-5s & 6-10s \\
Subscription &\$\satToUSD{6320}\result & \$\satToUSD{6320}\result & - & 3-5s & 3-5s & 6-10s\\
\midrule
HideMyAss & \$6.99 & $\infty$ & 60--180 min. & $\infty$ & $\infty$ & $\infty$ \\
PureVPN (VPN)& \$4.91 & $\infty$ & 2--60 min. & $\infty$ & $\infty$ & $\infty$ \\
\bottomrule
\end{tabular}}}
\label{tables:tithonus:download:time}
\vspace{-15pt}
\end{table}

\noindent
{\bf Tithonus vs. VPN costs}.
We now evaluate Tithonus' costs per expected request latency against those of a
VPN.
Table~\ref{tables:tithonus:download:time} summarizes our comparison. The top
rows show the costs and latencies of the pay-per access and subscription modes
of Tithonus.  These costs assume the current median transaction fee rate
according to \cite{bitcoinfees} for unobservability.
In $\S$~\ref{sec:evaluation:swift} we showed that increasing the fee rate for out-to-in communication has
no effect on the speed at which swift transactions are propagated through the
Bitcoin network. Further, for semi-interactive/concurrent communication (e.g.,
on-demand pay-per-access), the user is expected to be online at the time of the
request. Thus, for this use case, there is really no need to wait for
transactions to appear in the blockchain (see Table \ref{tables:tithonus:txncosts} in the appendix). In fact, we have also shown
($\S$~\ref{sec:evaluation:swift}) that having to resort to the blockchain
because of a missing swift transaction is a low probability event (0.0132 in
China).

To compare against VPN costs, we have not considered services that claim to
provide only directional, i.e., in-to-out, communications (e.g., NordVPN,
ViperVPN and ExpressVPN): since we are not in a censored country, we cannot
verify their claims. Instead, we have focused on two popular VPN providers,
HideMyAss and PureVPN, that publicize dual communication services (both
in-to-out and out-to-in) for China, similar to Tithonus. The monthly costs for
HideMyAss \cite{hmapricing} and PureVPN \cite{purevpnpricing} services at
publication time are \$6.99 and \$4.91 respectively. Communication setup
(payment verification and account activation) requires 60--90 min for PureVPN
and 2--60 min for HideMyAss.

Surprisingly, we found that despite the service being openly publicized on
their webpage, the access from within China and from the outside into China was
effectively blocked. Customer support blamed the  Great Firewall of China (GFW)
and was unable to provide an expected resolution time. They in fact recommended
asking for a refund which led us to believe this is not a temporary problem.
Thus, the expected latency for using their service is $\infty$, and, since the
number of requests per month that are available under any service plan is $0$,
the cost per request turns out to also be $\infty$, see
Table~\ref{tables:tithonus:download:time}.

\vspace{-5pt}

\subsection{Tithonus Client Computing Overhead}
\label{sec:evaluation:client}

\vspace{-5pt}

We used a 32-bit system with an Intel\textsuperscript{\tiny\textregistered}
Xeon\textsuperscript{\tiny\textregistered} Gold 6126 CPU @ 2.60GHz with 8GB RAM
to estimate the Tithonus client overhead. The time to retrieve the Tithonus
root certificate from the 200GB blockchain, using simple string matching, was
15m.36sec. When equipped with the Bitcoin client and the specific
transaction id, this time however becomes 4ms.

Depending on key sizes, in the on-demand, pay-per-access content fetch
solution, a client can process 38.8--57.26 thousand Tithonus txn/sec with a CPU
utilization of about 0.05\%.  Similarly, client registration and content
request messages achieve a speed of 1.2--1.67 million Tithonus txn/sec. Further,
for the altruistic directory and subscription service, our system achieved
2,850 ECDSA signature verifications per second. For comparison, the maximum
transaction rate ever processed by the Bitcoin network is 20 txn/sec.

\vspace{-10pt}

\subsection{Tithonus Performance Comparisons}
\label{sec:evaluation:comparison:tor}

\vspace{-5pt}

We now compare Tithonus with state of the art Blochain writing and censorship
resistance solutions. 

\noindent
\textbf{Comparison with Catena}. 
In the experiment of $\S$ \ref{sec:evaluation:comparison}, the {\it writing
efficiency} of our staged transactions is 0.93 (ratio of data bytes to total
bytes). For the same file, Catena~\cite{TD17} requires 172 transactions of size
282 bytes and 1 transaction of size 246, whose total size is 48,750 bytes, for
a writing efficiency of 0.28. At their recommended rate of 70 satoshi/byte, the
total cost of writing with Catena is 3,412,500 satoshi, or
\$\satToUSD{3412500}\result. Thus, Tithonus reduces the cost of sending
information through the Bitcoin blockchain, by 2 orders of magnitude (231 times
cheaper), and increases the writing efficiency by a factor of 3.

Further, the time for all Catena transactions to reach the blockchain amounts
to around 1,730 minutes (28.8 hours), as a transaction can be written only
after all the previous ones have been confirmed in the blockchain. The 9
Tithonus transactions took only 3.3s to be relayed by 1/8 of the good nodes in
China, and they were all mined in the blockchain in under 1.8 hours. Thus,
Tithonus improves on the transmission speed of a 14KB file by between 3 to 5
orders of magnitude (961 to 34,600 times faster). Tithonus achieves a goodput
of 4,601 B/s compared with Catena's 0.13 B/s.

\noindent
{\bf Comparison with Tor}. 
We deployed a VM with a paid VPN service that
tunnels all traffic from our lab in the US to the Shanghai province in China.
This infrastructure simulates a real life censored user. Since the Tor download
page is censored, we assume that the Tor user found an acceptable and secure
alternative way to download Tor, e.g., using Tithonus.

We tried all available Tor transports (obfs3, obfs4, meek, fte, scramblesuit)
and only meek-azure was able to bootstrap enough relays to establish a circuit
connection. In our experiments, this bootstrapping process lasted around 1.3
hours. In comparison, the communication setup for the \textit{pay-per-access}
mode in Tithonus requires 10--30 min (1--3 confirmations) until the first
cryptocurrency exchange credits the Tithonus account
\cite{BitfinexConfirmationTimes}. The meek pluggable transport uses the
infrastructure of large third parties to inflict collateral damage to
censorship attempts. Such \textit{domain fronting}-based solutions depend on
centralized third party collaboration, thus are not reliable (as recently
evidenced by Google suddenly disabling its support \cite{googleFronting}).  A
Tor user may obtain this bootstrapping information through Tithonus.

Subsequently, we have used Tor to download the Tithonus logo (13,804 Bytes) several times
using different circuits on each trial, and obtained download times ranging
between 10 to 15s. The VPN overhead, which we measured using the ping tool with
a payload of 13408 Bytes, was only 192 ms. Thus, Tor's 10-15 sec result suggests a 5
fold performance decrease when compared to Tithonus.

\noindent
{\bf Comparison with Collage}. 
Collage~\cite{BFV10} is a CRS that uses sites that host user-generated content
to communicate hidden messages. A user embeds messages into cover
traffic and posts them as content on a site.
Unlike Tithonus, Collage communications are free.  However, the latency and
goodput of Tithonus are better: Collage required 9 jpeg photos to store 23 KB
of data and took 1 minute to retrieve them. In contrast Tithonus achieves a
93\% storage efficiency and 3-5 seconds to retrieve (from China!) a comparable
amount of bytes.\\

\vspace{-25pt}

\subsection{Non-Conformant Nodes}
\label{sec:evaluation:analysis}

\vspace{-5pt}

\begin{figure}
\centering
\includegraphics[width=0.95\columnwidth]{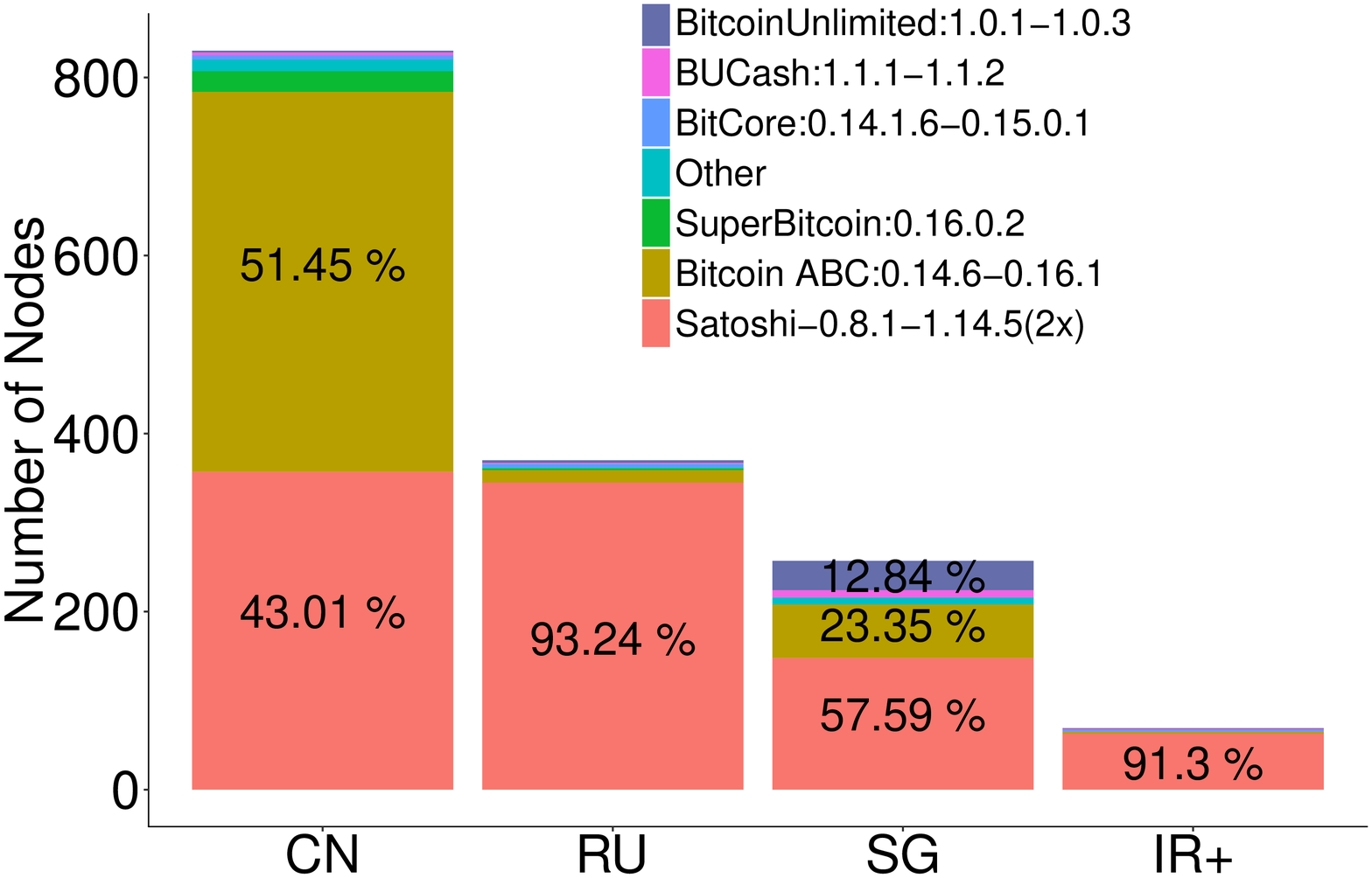}
\caption{Distributions of Bitcoin node client types in 12 countries with least
freedom of press. The only client expected to relay Tithonus transaction is the
Satoshi client (Bitcoin Core fork). China has the lowest percentage of such
clients, but still higher in absolute value than any of the other countries.}
\label{fig:client:distribution}
\vspace{-15pt}
\end{figure}

To understand why in the above experiments, only a fraction of  
Bitcoin nodes relayed our transactions, we have studied the advertised banners of the 1,526 contacted nodes. We have observed 67 total unique banners,
which we grouped into 7 categories, including an ``Other'' category that
contains banners with small representation (under 10 nodes).

The only nodes that can be expected to relay Tithonus transactions are the
Bitcoin Core (Satoshi)-compatible clients. The rest of
the clients, such as, Bitcoin Cash, Bitcoin Gold, etc.
have incompatible and independent blockchains.  Thus, we consider all other
nodes to be {\it non-conformant}, as they compete with the Satoshi nodes: they
accept connections from Satoshi clients but do not relay Bitcoin nor Tithonus
transactions. A Tithonus node that peers with such non-conformant
(Bitcoin forking) nodes may be more vulnerable to c-node filtering attacks, and
even eclipse attacks. Specifically, non-conformant nodes affect not only
Tithonus but also the distribution of regular Bitcoin Core transactions. This
occurs because Bitcoin Core nodes waste outgoing connection slots (only 8
available by default) when peering with nodes that effectively reduce their
reachability and connectivity to the Bitcoin network.  However, we note that
even in the presence of a large percentage of non-conformant nodes, Tithonus
transactions were relayed by a large number (99.24\%) of conformant nodes in
censored countries.

Figure~\ref{fig:client:distribution} shows the distribution of the advertised
Bitcoin node types for the 1,526 nodes, over the top 3 censored countries in
terms of node population and the remaining 9 as a whole. China has the highest
percentage of non-conformant nodes, followed by Singapore. China is dominated by
the Bitcoin Cash (ABC) client. The number of Satoshi nodes in China
exceeds those in Russia, where they form the vast majority of nodes.

\vspace{-10pt}

\section{Discussion and Limitations}
\label{sec:discussion}

\vspace{-5pt}

\noindent

\noindent
{\bf Limitations of Tithonus}.
Tithonus is expensive, unsuitable for low-latency interactive communications,
and limits the amount of content that can be requested per time unit.  Tithonus
is suitable when other solutions are blocked (e.g.,
VPNs~\cite{VPN.China.a,VPN.China.b,VPN.Russia}) or their detectability may
expose the user. Tithonus can be used to bootstrap other censorship-resistant
systems, e.g., to download their source code or communicate Tor bridge IPs.

Tithonus communications are stored in the blockchain, where they can be
accessed and decrypted by an adversary at a later time. However, Tithonus
messages do not include client information. To attribute such messages to the
source clients, an adversary would also need to monitor and record client
communications, in which case the blockchain provides no advantage.

\noindent
\textbf{Payment obfuscation.} The content fetch solutions described in $\S$~\ref{sec:tithonus:fetch} require the client to transfer funds to an
address, then enable the Tithonus service to redeem those funds. However, the
server cannot just transfer the funds to a Tithonus controlled address or
immediately spend the funds on an identifiable resource (e.g., through a
transaction that uses as inputs multiple such addresses). Such actions would
inform the censor about the identities of clients who use Tithonus. To
address this vulnerability, the Tithonus server makes use of cryptocurrency
exchanges that break the connection between the coins spent and the resource
obtained. Since some of these exchanges could be in collusion or have been
compromised by the censor, the Tithonus server needs to chain the use of
several exchanges before performing the final use of the funds. Specifically,
the funds are transferred to one exchange and traded to several different
cryptocurrencies before being transferred to another exchange. Depending on the
desired confirmation speed, transferring funds to an exchange currently incurs
a Bitcoin transaction fee of 200-600 satoshis. In addition, trading operations
have a cost of around 2\% of the transaction nominal value.

\noindent
{\bf Censorship resilience}.
The cost of blocking Bitcoin may be too large even for the most powerful
censors. Censors may have invested heavily in
cryptocurrencies~\cite{ChinaMines} or have mining advantages over
competitors~\cite{asicBoost}, thus may be reluctant to block cryptocurrencies.
A censor who is unwilling to completely block the use of Bitcoin will be unable
to prevent out-to-in communications from reaching the blockchain, thus censored
clients (see $\S$~\ref{sec:evaluation:comparison}).

Censoring in-to-out transactions suspected of being part of Tithonus can split the
blockchain, as it will lead to inconsistent mempools and different blocks for
inside and outside miners~\cite{decker2013information}. The inherent false
positive rate of any filtering algorithm will also lead to the censor blocking
valid transactions. Tithonus seeks to increase this FPR, by making its
transactions blend in Bitcoin.  We also found evidence that the default Bitcoin
node policy to select peers actively resists such attacks. In
Section~\ref{sec:evaluation:swift} we show that even when 70\% of the nodes in
the censored area are non-conformant, Tithonus transactions reach most benign
nodes in less than 19 seconds.

Further, we note that an adversary could exploit knowledge that the censor
blocks Tithonus transactions (even if he allows their hashes) to deploy double
spending attacks against users within the censored area. For instance, an
adversary outside the censored area could issue a transaction to pay an ``outside''
service, that mimics Tithonus staged transactions. The censor would then prevent this
transaction from reaching services in the censored area. The adversary then
double spends the input address of this staged transaction, by issuing a
legitimate (no Tithonus payload) transaction from it that pays a service within
the censored area. Since the service did not see the first transaction, it will
accept the second one.

\noindent
{\bf Middleman blockchain attack}.
In an effort to remedy the shortcomings of the previous attack, 
the censor could attempt forcing censored users to make use of a censor-controlled \textit{blockchain accessing service}. The goal of this attack would be to prevent
censored nodes from directly accessing an uncensored version of the blockchain. Instead, censored nodes
would need to contact this blockchain accessing service, ``the middleman'', to fetch blockchain information and perform transaction and block
verifications. 
The middleman
would then be able to, e.g., prevent access to the blocks containing
Tithonus
data but still answer queries about the state of the Bitcoin network.

Such a service would however make participants inside the censored area
vulnerable to attacks, e.g., double spending with merchants within the censored
region.
Specifically, attackers could attempt to exploit the inherent racing condition
that arises from the time the middleman nodes receive a to-be censored
transaction and a query from a user inside the censored region. In addition,
the middleman service nodes would have to be publicly and easily identifiable
for censored users to use them. This visibility makes them easy targets to
attacks, e.g., DoS, selective malicious information feeding and even total
eclipsing.

Bitcoin provides resilience to such attacks by incentivizing a diverse
and distributed ecosystem of nodes. The middleman
blockchain attack reduces this diversity and thus reduces the security of its
participants, as studied by Decker and
Wattenhofer~\cite{decker2013information}.

\textbf{Malleability concerns.} Since its creation, the Bitcoin ecosystem has
struggled with the problem of transaction malleability. Thus, Tithonus
transactions are also malleable. This vulnerability could pose problems for
staged transactions that require integrity protections when sent
simultaneously. However, in order for a censor to perform a ``malleability +
rushing'' attack on Tithonus, it needs to win a race against the
rest of the honest Bitcoin network and prevent un-tainted Tithonus-issued
writing transactions from reaching their intended clients. Such a DoS attack
could be easier achieved by direct filtering. However, such actions would harm
the Bitcoin ecosystem and miners inside the censored
area~\cite{decker2013information}, placing this attack outside of our threat
model.

\vspace{-15pt}

\section{Related Work}
\label{sec:related}

\vspace{-5pt}

\noindent
{\bf Blockchain based censorship resistance}.
Wachs et al.~\cite{wachs2014censorship} have evaluated the feasibility of
building a censorship resilient, privacy preserving domain name system.
Tomescu and Devadas~\cite{TD17} proposed Catena, a blockchain-based
non-equivocation solution. Catena uses transaction chains where each
transaction has two outputs, one that stores data into an unspendable
OP\_RETURN output and one that is spent in the following transaction in the
chain. Catena transactions are easy to fingerprint by the censor and have lower
goodput and a higher price than Tithonus ($\S$~\ref{sec:evaluation}).

\noindent
{\bf Proxy based and decoy routing censorship resistance}.
Tithonus imposes higher costs and latency than existing VPN services. However,
VPNs are easy to block, with countries like
China~\cite{VPN.China.a,VPN.China.b} and Russia~\cite{VPN.Russia} curtailing
access to VPNs.  Decoy routing deploys relay stations to routers of
participating ISPs and leverages covert channels (see below) to hide
information inside requests to an overt destination, which is then detected and
processed by a relay station.  Decoy routing leverages collateral damage
assumptions (see next) to prevent censorship. Many decoy routing systems are
vulnerable to latency analysis and website fingerprinting attacks. Bocovich and
Goldberg~\cite{BG16} addressed this problem through a decoy routing solution
that mimics access to an uncensored site.

Both proxy based solutions and decoy routing use intermediate participants to
route traffic. Similar to Tithonus, intermediaries increase communication
latency.  Further, both decoy routing and proxy based CRS assume voluntary
participation of multiple participants.  Tithonus provides financial incentives
for participation. Tithonus can be used to distribute IPs of Tor bridges, but
does not prevent a censor from discovering and blocking them.

\noindent
{\bf High collateral damage CRS}.
Emerging solutions attempt to bypass such attacks, by leveraging the
unwillingness of censors to block access to infrastructure providing large
scale access to benign resources. For instance, Holowczak and
Houmansadr~\cite{HH15} found that the Chinese firewall does not block IPs of
CDN providers that store censored content, as they also store large amounts of
benign content. However, Zolfaghari and Houmansadr~\cite{ZH16} found that CDN
based CRS (e.g., CacheBrowser~\cite{HH15}) can leak the identity of destination
websites and are vulnerable to website fingerprinting attacks. They designed
CDNReaper, a CDN-aware based CRS that addresses these attacks, e.g., by
processing the requested censored content.

Fifield et al.~\cite{FLHWP15} further proposed domain fronting, that sets up
circumvention proxies on web services that share IP addresses with other benign
services.  While blocking all such IPs (including CDN IPs) is possible to a
powerful censor, it would block access to content considered benign and even
useful to the censor.  

\noindent
{\bf Mimicry and tunneling based CRS}.
Mohajeri et al.~\cite{MLDG12} proposed SkypeMorph, a mimicry based CRS that
morphs Tor traffic to resemble the characteristics of Skype calls.  Houmansadr
et al.~\cite{HRBS13} introduced FreeWave, a CRS that modulates censored traffic
into acoustic signals which it tunnels over VoIP (i.e., Skype) connections.
Unlike Tithonus, SkypeMorph and FreeWave communications are free.
However, payments provide incentives for running the Tithonus
service, and resilience against DoS attacks.
Tor can be blocked even when using SkypeMorph, since a censor impersonating
valid users can discover and block Tor bridges. SkypeMorph is vulnerable to
packet drop attacks. In contrast, filtering attacks do not impact Tithonus when
the censor does not want to affect Bitcoin usage.

\vspace{-15pt}

\section{Conclusions}

\vspace{-5pt}

We introduced Tithonus, a new CRS built on the Bitcoin network and blockchain.
We develop solutions for Tithonus clients to fetch censored data of arbitrary
size, that are 2 orders of magnitude cheaper and 3-5 orders of magnitude faster
than state of the art Bitcoin writing solutions. Tithonus is robust even in the
presence of non-conformant nodes, and this robustness is not affected by the
use of low fee transactions, in the absence of congestion. Thus, Tithonus is
able to provide an optimally cheap solution within a given cryptocurrency
ecosystem.  Tithonus is practical when considering its reach of Bitcoin nodes
available in censored countries.

\vspace{-15pt}

\section{Acknowledgments}

\vspace{-5pt}

We thank the shepherd and the anonymous reviewers for their excellent feedback.
This research was supported by NSF grant CNS-1526494.

\vspace{-15pt}

\bibliographystyle{unsrt}
\bibliography{main}

\appendix

\section{Proof Sketch of Indistinguishability of In-To-Out Requests}
\label{appendix:proof}

We introduce the following indistinguishability game, played by a
challenger and an adversary.  First, the adversary chooses a message $M$ and
sends it to the challenger.  The challenger encrypts $M$ and takes the leftmost
28 bytes of the result to generate $D$. He then randomly picks a bit $b$. If
$b=0$, the challenger generates and outputs a regular compressed public key. If
$b=1$, he embeds $D$ in a public key as described in Section
\ref{sec:tithonus:unit} and outputs the embedding instead. The adversary then
outputs a guess $b'$ for the challenger's bit $b$. The adversary wins if $b'$ =
$b$ with probability non-negligibly higher than 1/2.

We observe that to prove that the Tithonus multisig p2sh constructs are
indistinguishable from regular multisig p2sh transactions, it suffices to show
that our procedure for embedding encrypted data ($\S$~\ref{sec:tithonus:unit})
produces outputs that are indistinguishable from a compressed elliptic curve
public key.  This sufficiency claim follows from the fact that a our p2sh
construct simply consists of two of these embeddings and a real public key,
whereas a regular p2sh Bitcoin transaction consists of 3 real public keys.
Thus, if each of these embeddings are indistinguishable from public keys, our
construct and regular p2sh transactions must be indistinguishable as well.

To see that our embeddings are indistinguishable from a compressed elliptic
curve public key, we observe that {\it by definition}, a compressed public
key consists of the prefix 0x02/0x03 followed by any random number $\tilde{x}$
that satisfies the following conditions $\mathcal{C}$:

\begin{compactitem}
        \item $\tilde{x}$ is smaller than 0xFF...EFFFFFC2F (the 32 byte prime $p$ used in Bitcoin's sec256pk1),
        \item $\tilde{x}$ satisfies $\tilde{w} = \tilde{x}^3 + 7$ (the elliptic curve equation used in Bitcoin's sec256pk1), where:
        \item $\tilde{w}$ is a quadratic residue (QR) in $\mathbb{F}_p$.
\end{compactitem}

In addition, {\it by construction}, our embedding consists of the prefix
0x02/0x03, followed by $x=D,R$ such that:

\begin{compactitem}
        \item $D$ is smaller than 0xFF...FF ($2^{224}-1$) and $R$ is smaller than 0xFFFFFC2F, so that $x$ is smaller than the prime $p$,
        \item $x$ also satisfies $w = x^3 + 7$, where:
        \item $w$ is also a QR in $\mathbb{F}_p$.
\end{compactitem}

Thus, to establish indistinguishability all we need to show is that $x=D,R$ is
indistinguishable from a random number $\tilde{x}$ that satisfies conditions
$\mathcal{C}$. To this end, we notice that, since $R$ is a random number, if
an adversary is able to differentiate between $x$ and $\tilde{x}$, then
she would also be able to differentiate between $D$ and a random
number. Such an adversary would then also have a non-negligible advantage in
differentiating the output of the Rjindael algorithm and a random number.
Although there is in fact no proof that Rjindael is indeed a secure PRF, this
is a generalized assumption about the Rjindael cipher~\cite{daemen1999aes}.

Finally, we observe that unspent keys in Tithonus multisig p2sh transactions
are not suspicious: Bitcoin recommends that users do not reuse addresses, to
prevent linkability attacks.

\section{Tithonus Fees}

\begin{table}[!ht]
\centering
\small
\caption{
\bf{Tithonus economic costs for different fee rates. The current median transaction fee is 9 Sat/byte. We emphasize (bold) the recommended fees. The rest of the fee values are not recommended since they provide no advantage to Tithonus users. Monetary costs are displayed at current Bitcoin price (1 BTC
$\approx$ \$\usd{\btcPrice}).}
}
\vspace{-5pt}
\resizebox{0.49\textwidth}{!}{
\textsf{
\begin{tabular}{lccc c}
\toprule
Tithonus txn type & 1 Sat/byte & 4 Sat/byte & 9 sat/byte  & 16 sat/byte \\ 
\midrule
Out-to-in (14KB file) & \textbf{ \$\satToUSD{14722}\result }& \$\satToUSD{58888}\result & \$\satToUSD{132498}\result & \$\satToUSD{235552}\result \\ 
Avg. News Article Cost& \textbf{ \$\satToUSD{15408}\result} & \$\satToUSD{61632}\result & \$\satToUSD{138672}\result & \$\satToUSD{123264}\result \\ 
In-to-out (Reg) & \$\satToUSD{1580}\result& \$\satToUSD{6320}\result & \textbf{\$\satToUSD{14220}\result} &  \$\satToUSD{25280}\result \\ 
In-to-out (Request)& \$\satToUSD{1580}\result& \$\satToUSD{6320}\result & \textbf{\$\satToUSD{14220}\result} &  \$\satToUSD{25280}\result \\ 
\bottomrule
\end{tabular}}}
\label{tables:tithonus:txncosts}
\vspace{-15pt}
\end{table}

The calculations presented in Section~\ref{sec:evaluation:comparison} select
the minimum fee rate that achieves the fastest on-chain time according to
\cite{bitcoinfees} (currently at \btcFee~sat/byte).
Table~\ref{tables:tithonus:txncosts} provides details for Tithonus transaction
costs for different Bitcoin fee rates, including the minimum, the current Bitcoin median
transaction fee (9 sat/byte), and other fees that are not recommended and that provide
no advantage to users.

\end{document}